\documentclass[aps,amsmath,latexsym,amssymb,prl,onecolumn,graphicx,epsfig]
{revtex4}

\usepackage{epsfig}

\newcommand{\bkp}{\mbox{\boldmath $k_{\parallel}$}}

\newcommand{\sbkp}{\mbox{\footnotesize \boldmath $k_{\parallel}$}}

\newcommand{\bTpar}{\mbox{\boldmath $T_{\parallel}$}}
\newcommand{\bTper}{\mbox{\boldmath $T_{\perp}$}}
\newcommand{\Tpar}{\mbox{ $T_{\parallel}$}}
\newcommand{\Tper}{\mbox{ $T_{\perp}$}}

\newcommand{\bS}{\mbox{\boldmath $S$}}
\newcommand{\bH}{\mbox{\boldmath $H$}}
\newcommand{\be}{\mbox{\boldmath $e$}}
\newcommand{\bsig}{\mbox{\boldmath $\sigma$}}
\newcommand{\bj}{\mbox{\boldmath $j$}}

\newcommand{\bDel}{\mbox{\boldmath $\Delta$}}
\newcommand{\bT}{\mbox{\boldmath $T$}}
\newcommand{\bP}{\mbox{\boldmath $P$}}
\newcommand{\bp}{\mbox{\boldmath $p$}}
\newcommand{\bm}{\mbox{\boldmath $m$}}
\newcommand{\bG}{\mbox{\boldmath $\Gamma$}}
\newcommand{\sa}{\mbox{$\sin^{2} \alpha _{0}$}}
\newcommand{\ca}{\mbox{$\cos^{2} \alpha _{0}$}}
\newcommand{\sph}{\mbox{$\sin^{2} \phi _{0}$}}

\begin{document}
\title{A selfconsistent theory of current-induced switching of magnetization}
\author{D.M. Edwards$^{1}$, F. Federici$^{1,2}$, J. Mathon$^{2}$, and A.Umerski$^{3}$}
\address{$^{1}$Department of Mathematics, Imperial College, London SW7 2BZ, U.K.,
$^{2}$Department of Mathematics, City University, London EC1V 0HB, U.K.,
$^{3}$Department of Applied Mathematics, Open University, Milton Keynes MK7 6AA, U.K.}
\date{\today}
\begin{abstract}
A selfconsistent theory of the current-induced switching of magnetization using nonequilibrium Keldysh formalism is developed for a junction of two ferromagnets separated by a nonmagnetic spacer in the ballistic limit. It is shown that the spin-transfer torques responsible for current-induced switching of magnetization can be calculated from first principles in a steady state when the magnetization of the switching magnet is stationary. A steady state is achieved when the spin-transfer torque, proportional to bias voltage in the linear response regime, is balanced by the torque due to anisotropy fields. The spin-transfer torque is expressed in terms of one-electron surface Green functions for the junction cut into two independent parts by a cleavage plane immediately to the left and right  of the switching magnet. The surface Green functions are calculated using a tight-binding Hamiltonian with parameters determined from a fit to an {\it ab initio} band structure. This treatment yields the spin transfer torques taking into account rigorously contributions from all the parts of the junction.
The spin-transfer torque has two components, one with the torque vector $\bTpar $ in the plane containing the magnetizations of the two magnetic layers and another with the torque vector $\bTper $ perpendicular to this plane. It is shown that, in general, $\bTpar $ and $\bTper$ may be comparable in magnitude and they both tend to finite values independent of the spacer thickness in the limit of a thick spacer. $\bTper $ is shown to be small when the exchange splitting of the majority- and minority-spin bands in both ferromagnets tends to infinity or in the case when the junction has a plane of reflection symmetry at the center of the spacer. The torques $T_{\perp} $ and $T_{\parallel} $ are comparable for a Co/Cu/Co(111) junction when the switching Co layer is one or two atomic planes thick. $T_{\perp} $ is $\approx$ 27\% of $T_{\parallel} $ even for a switching Co magnet of  ten atomic planes. Depending on material parameters of the junction, the relative sign of $T_{\perp}$ and $T_{\parallel}$ can be negative as well as positive. In particular, $T_{\perp}/T_{\parallel}<0$ for Co/Cu/Co(111) with switching Co magnet of one atomic plane and $T_{\perp}/T_{\parallel}>0$ for two atomic planes of Co. A negative sign of the ratio $T_{\perp}/T_{\parallel}$ has a profound effect on the nature of switching, particularly in the realistic case of easy-plane (shape) anisotropy much larger than in-plane uniaxial anisotropy. To calculate the hysteresis loops of resistance versus current, and hence to determine the critical current for switching, the microscopically calculated spin-transfer torques are used as an input into the phenomenological Landau-Lifshitz equation with Gilbert damping.
In the absence of an applied magnetic field, an ordinary hysteresis loop is the only possible switching scenario when $T_{\perp}/T_{\parallel}>0$. However, for $T_{\perp}/T_{\parallel}<0$, a normal hysteretic switching occurs only at relatively low current densities. When the current exceeds a critical value, there are no stable steady states and the system thus remains permanently in a time dependent state. This is analogous to the observed precession of the switching magnet magnetization caused by a DC current in the presence of an applied magnetic field. 
The present calculations for Co/Cu/Co(111) show that the critical current for switching in the hysteretic regime is $\approx 10^{7}A/cm^{2}$, which is in good agreement with experiment.

\end{abstract}

\maketitle

\noindent
\section{Introduction}

\noindent
Slonczewski \cite{slon} proposed a new method of switching the magnetization direction of a thin film by means of a spin-polarized current. The current is spin-polarized by passing through a thick layer of a ferromagnetic metal, whose magnetization is assumed to be pinned, subsequently passing through a nonmagnetic metallic spacer layer and then through a thin magnetic switching layer into a nonmagnetic lead.
Early related theoretical work is due to Berger \cite{berger}. Switching of the magnetization is accompanied by a change in resistance (CPP GMR) and the effect has been observed experimentally by studying hysteresis loops in resistance versus current plots for pillar systems \cite{pillar}. Jumps in the hysteresis curve occur between steady states of constant current and static magnetization, just as in the Stoner-Wohlfarth \cite{stoner} theory of field-switching jumps occur between equilibrium states. We have formulated a first-principle theory of current-induced switching based on this idea. As in the Stoner-Wohlfarth theory, we assume that the switching magnet remains single domain during the switching process. 

One of the main aims of this paper is to calculate hysteresis loops of resistance versus current from first principles for a much more general situation than has been considered previously. In previous treatments a uniaxial anisotropy field was introduced in the switching magnet with its direction parallel to the magnetization of the polarizing magnet \cite{sun}, \cite{slon}. In this case, there are only two steady states in which the magnetizations of the polarizing and switching magnets are either parallel or antiparallel. It is for this reason that the steady-state approach has not previously been further developed.  However, in real experiments on pillar structures shape anisotropy due to variable shape of pillar cross sections means that the direction of the anisotropy field in the switching magnet is not simply related to  the direction of the magnetization of the polarizing magnet. Furthermore, in some experiments \cite{kiselev}, \cite{grollier} an external field is also applied so that more general orientations of the magnetizations occur in the steady states. It is, therefore, essential to consider a completely general case when the uniaxial anisotropy field makes an arbitrary angle $\theta $ with the polarizing magnetization. We also include  the easy plane anisotropy, which is always large in layered magnets, and investigate fully its consequences. 

In a steady state there is a balance between the spin-current torque, acting on the switching magnet due to the spin-polarized current, and the torque due to anisotropy and external fields. In our general first-principle treatment, two components of the spin-current torque appear naturally, one with the torque vector $\bTpar $ in the plane containing the magnetizations of the two magnetic layers ('in-plane' torque) and another with the torque vector $\bTper $ perpendicular to this plane ('out-of-plane' torque). Slonczewski \cite{slon} considered only $\bTpar $ and it is generally believed \cite{stiles} that $\bTper $ is negligible. We start the presentation of our results in Sec.6 by deriving from the general Keldysh formalism the results of Slonczewski's original calculations \cite{slon}, in which only $\bTpar $ appears. It will be seen that this result is not always  valid  but is just an artefact of Slonczewski's simple model. In fact, we shall show that in some cases
$\bTper $ is dominant and that, even when small, $\bTper $ is essential since its importance is strongly enhanced in the presence of easy-plane anisotropy.  

To calculate hysteresis loops for this general scenario, we need to solve the following problems:
(i) calculate microscopically both the in-plane and out-of-plane components of the spin-current torque;
(ii)  determine the steady states which form the continuous parts of the hysteresis curve;
(iii)  investigate the stability of such states in order to determine critical currents at which the jumps, and hence switching, occur;
(iv) calculate the resistance of the layered structure along the steady-state paths.

Within our unified theory, all this can be done for a general layered system with a fully realistic band structure.

A jump in the hysteresis curve occurs at a critical current when one steady state becomes unstable and the system seeks out another stable steady state. This is in analogy with the Stoner-Wohlfarth \cite{stoner} theory of field-switching where one deals with equilibrium states instead of the present nonequilibrium steady states. As in that theory we do not concern ourselves with the detailed dynamics of the switching. However, we identify in this paper certain cases in which one steady state becomes unstable above a critical current but there are no other stable steady states available. Under these circumstances the magnetization of the switching layer remains perpetually in a time-dependent state.

In order to study nonequilibrium steady states we use the Keldysh formalism \cite{keldysh}, \cite{caroli}, \cite{edwards}  described in Sec. 3. As pointed out above, the steady state arises from a balance between spin-current torque and anisotropy field torque. Hence it is essential to include anisotropy and/or external fields in the Hamiltonian of the system from the outset. It is also necessary to treat correctly the on-site electron-electron interaction which is responsible for the spontaneous magnetization of the polarizing and switching magnets. This is achieved by insisting that the local exchange field is in the direction of the local magnetization, which is the essential feature of self-consistent field approximations such as unrestricted Hartree-Fock (HF) and local spin density (LSDA). Such a treatment respects the spin-rotational symmetry of the ferromagnet in the absence of external fields. Beyond this we do not need to introduce a self-consistent treatment of the Coulomb interaction explicitly, although bulk LSDA calculations underlie the band parameters and exchange splittings used in our calculations. In our Keldysh approach the direction of magnetization in each atomic plane of the switching magnet is determined self-consistently in the steady state by the requirement
that the magnetization of a given atomic plane is parallel to the exchange field in that plane. The relationship between this approach and the more intuitive one of balancing torques is discussed in Secs. 2 and 5.

The treatment described above enables us to determine all possible steady states of the system and the next step is to investigate their stability. We do this by introducing the spin-current torques, calculated microscopically as functions of magnetization direction, and anisotropy torques into a Landau-Lifshitz equation of motion for the magnetization including Gilbert damping. We linearize the equation of motion about the steady-state solution to obtain the conditions for stability.

Finally, we construct hysteresis curves from continuous steady-state paths and jumps at points of instability.
\\

\noindent
\section {Theoretical model}

\noindent
The layer structure we consider is shown in Fig.1. It consists of a

\begin{figure}[here]
\centerline{\epsfig{figure=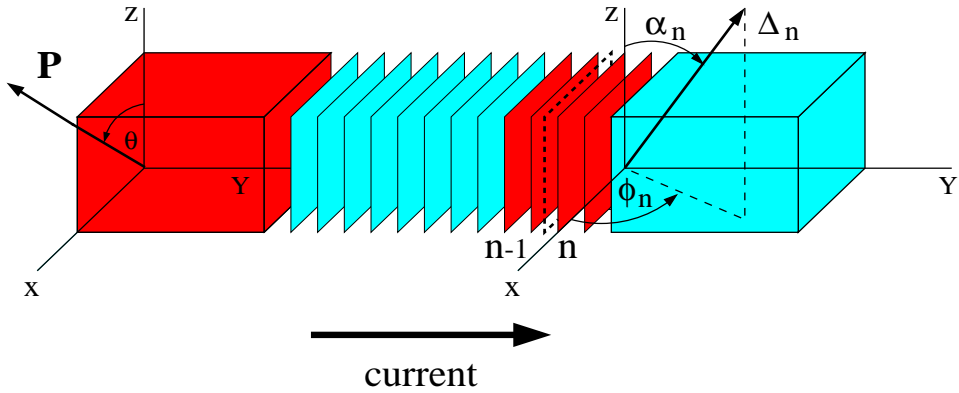,width=5in,height=2in,angle=0}}
\caption{\footnotesize Schematic picture of a magnetic layer structure for current-induced switching (magnetic layers are darker, non-magnetic layers lighter).} 
\end{figure}
\noindent
semi-infinite polarizing ferromagnet with magnetization $\bP$, a nonmagnetic metallic spacer with N atomic planes, a switching magnet with M atomic planes, and a semi-infinite nonmagnetic lead of the same material as the spacer. Each layer is described by a tight-binding model, in general multiorbital with s, p, and d orbitals whose one-electron parameters are fitted to first-principle bulk band structure, as discussed previously \cite{prb97}. The Hamiltonian is, therefore, of the form

\begin{eqnarray}
H=H_{0} + H_{int} + H_{anis},
\label{eq1}
\end{eqnarray}
\noindent
where the one-electron hopping term $H_{0}$ is given by

\begin{eqnarray}
H_{0} = \sum_{\sbkp \sigma} \sum_{m \mu ,n \nu}T_{m\mu ,n\nu}(\bkp)
c^{\dagger}_{\sbkp m \mu \sigma}c_{\sbkp n \nu \sigma},
\label{eq2}
\end{eqnarray}
\noindent
where $c^{\dagger}_{\sbkp m \mu \sigma}$ creates an electron in a Bloch state, with in-plane wave vector $\bkp $ and spin $\sigma$, formed from a given atomic orbital $\mu $ in plane m. H$_{int}$ is an on-site interaction between electrons in d orbitals which leads to an exchange splitting of the bands in the ferromagnets and is neglected in the spacer and lead. The magnetization of the polarizing magnet is assumed to be pinned in the (z,x)-plane, making an angle $\theta $ with the z axis, as shown in Fig.1. $H_{anis}$ contains effective fields in the switching magnet corresponding to uniaxial $\bH_{u}$ and easy-plane $\bH_{p}$ anisotropies. It is given by
\begin{eqnarray}
H_{anis}=-\sum_{n}\bS_{n}.\bH_{A},
\label{eq3}
\end{eqnarray}
\noindent
where $\bS_{n}$ is the operator for the total spin angular momentum of plane $n$ and 
\begin{eqnarray}
\bH_{A}=\bH_{u}+\bH_{p}.
\label{eq4}
\end{eqnarray}
\noindent
$\bH_{u}$ and $\bH_{p}$ are given by
\begin{eqnarray}
\bH_{u}=(\be_{z}.<\overline {\bS_{n}}>)H_{u0}\be_{z}
\label{eq5}
\end{eqnarray}

\noindent
\begin{eqnarray}
\bH_{p}=-(\be_{y}.<\overline {\bS _{n}}>)H_{p0}\be_{y},
\label{eq6}
\end{eqnarray}
\noindent
where $<\overline {\bS_{n}}>$ is a unit vector in the direction of the thermal average of  $\bS_{n}$, and $\be_{x}$, $\be_{y}$, $\be_{z}$ are unit vectors in the direction of the axes shown in Fig.1. $H_{u0}$, $H_{p0}$ measure the strengths of the uniaxial and easy-plane anisotropies and have dimensions of frequency. These quantities may be converted to a field in tesla by multiplying them by $\hbar/2\mu_{B}=5.69\times10^{-12}$. We assume that anisotropy fields are uniform throughout the switching magnet but it would be easy to generalize to include, for example, a surface anisotropy. 
The spin angular momentum operator $\bS_{n}$ is given by
\begin{eqnarray}
\bS_{n}=\frac{1}{2}\hbar \sum_{\sbkp \mu}(c^{\dagger}_{\sbkp n \mu \uparrow}, c^{\dagger}_{\sbkp n \mu \downarrow})\bsig (c_{\sbkp n \mu \uparrow}, c_{\sbkp n \mu \downarrow})^{T}
\label{eq7}
\end{eqnarray}
\noindent
and the corresponding operator for spin angular momentum current between planes $n-1$ and $n$ is
\begin{eqnarray}
\bj_{n-1}=-\frac{i}{2}   \sum_{\sbkp \mu \nu}T(\bkp)_{n\nu,n-1\mu}(c^{\dagger}_{\sbkp n \nu \uparrow}, c^{\dagger}_{\sbkp n \nu \downarrow})\, \bsig (c_{\sbkp n-1 \mu \uparrow}, c_{\sbkp n-1 \mu \downarrow})^{T} + h.c.
\label{eq8}
\end{eqnarray}
\noindent
Here, $\bsig =(\sigma_{x},\sigma_{y},\sigma_{z})$, where the components are Pauli matrices and Eq.(8) yields the charge current operator if $\frac{1}{2} \bsig $ is replaced by a unit matrix multiplied by the electronic charge $e/\hbar$, where e is the electronic charge (negative).

All currents flow in the y direction, perpendicular to the layers, and the components of the vector $\bj $ correspond to transport of x, y, and z components of spin. The rate of change of $\bS _{n}$ in the switching magnet  is given by
\begin{eqnarray}
i\hbar \dot{\bS _{n}}=[\bS _{n},H_{0}]+[\bS _{n},H_{anis}]
\label{eq9}
\end{eqnarray}
\noindent
since the spin operator commutes with the interaction Hamiltonian $H_{int}$.

It is straightforward to show that
\begin{eqnarray}
[\bS _{n},H_{0}]=i\hbar (\bj _{n-1}-\bj _{n})
\label{eq10}
\end{eqnarray}
\noindent
and
\begin{eqnarray}
[\bS _{n},H_{anis}]=-i\hbar (\bH _{A}\times \bS _{n}),
\label{eq11}
\end{eqnarray}
\noindent
In a steady state, the magnetization is time-independent so that 
$<\dot{\bS _{n}}>=0$. Hence
\begin{eqnarray}
<\bj _{n-1}> - <\bj _{n}> = \bH _{A}\times <\bS _{n}>.
\label{eq12}
\end{eqnarray}
\noindent
The left-hand side of Eq.(12) corresponds to the rate of transfer of spin angular momentum to plane $n$ in the steady state. Thus Eq.(12) shows explicitly how, in the steady state, this spin-transfer torque is balanced by the torque due to anisotropy fields. The concept of spin-transfer torque was first introduced by Slonczewski \cite{slon}.

\section{Keldysh formalism}

In this section we show how to calculate the spin current $<\bj _{n-1}>$ and spin density $<\bS _{n}>$ in the non-equilibrium steady state and verify that they are related by Eq.(12). To produce a spin-polarized current in the system we apply a bias $V_{b}$ between the polarizing magnet and the lead. To use the Keldysh formalism \cite{keldysh}, \cite{caroli}, \cite{edwards} to calculate $<\bj _{n-1}>$ and $<\bS _{n}>$ we consider an initial state at time $t=-\infty $ in which the hopping integral $T_{n\nu ,n-1\mu}$ between planes $n-1$ and $n$ is switched off. Then both sides of the system are in equilibrium but with different chemical potentials $\mu _{L}$ on the left and $\mu _{R}$ on the right, where $\mu _{L}-\mu _{R}=eV_{b}$. The interplane hopping is then turned on adiabatically and the system evolves to a steady state. The cleavage plane, across which the hopping is initially switched off, may be taken in either the spacer or switching layer or in the lead. Fig.1 shows the situation when the cleavage plane is between atomic planes $n-1$ and $n$ in the switching magnet. In principle, the Keldysh method is valid for arbitrary bias $V_{b}$ but here we restrict ourselves to small bias corresponding to linear response. This is reasonable since for larger bias electrons would be injected into the switching magnet far above the Fermi level and many-body processes neglected here would be important. Furthermore, in  metallic systems the bias will never be large.

Following Keldysh \cite{keldysh}, \cite{caroli}, we define a two-time matrix
\begin{eqnarray}
G^{+}_{RL}(t,t') = i<c^{\dagger }_{L}(t')c_{R}(t)>,
\label{eq13}
\end{eqnarray}
\noindent
where $R\equiv (n,\nu ,\sigma ')$ and $L\equiv (n-1,\mu ,\sigma )$, and we suppress the $\bkp $ label. The thermal average in Eq.(13) is calculated for the steady state of the coupled system. The matrix $G^{+}_{RL}$ has dimensions $2m\times 2m$ where $m$ is the number of orbitals on each atomic site, and is written so that the $m\times m$ upper diagonal block contains matrix elements between  $\uparrow $ spin orbitals and the $m\times m$ lower diagonal block relates to $\downarrow $ spin. $2m\times 2m$ hopping matrices $T_{LR}$ and $T_{RL}$ are written similarly and in this case only the diagonal blocks are nonzero. If we denote $T_{LR}$ by $T$, then $T_{RL}=T^{\dagger}$. We also generalize the definition of $\bsig $ so that its components are now direct products of the $2\times 2$ Pauli matrices $\sigma _{x}, \sigma _{y}, \sigma _{z}$ and the $m\times m$ unit matrix.
The thermal average of the spin current operator, given by Eq.(8), may now be expressed as 
\begin{eqnarray}
<\bj _{n-1}> = \frac {1}{2} \sum_{\sbkp} Tr \{[G^{+}_{RL}(t,t)T - G^{+}_{LR}(t,t)T^{\dagger}]\bsig\}.
\label{eq14}
\end{eqnarray}
\noindent
Introducing the Fourier transform $G^{+}(\omega)$ of $G^{+}(t,t')$, which is a function of $t-t'$, we have
\begin{eqnarray}
<\bj _{n-1}> = \frac {1}{2} \sum_{\sbkp} \int \frac{d\omega}{2\pi}Tr \{[G^{+}_{RL}(\omega)T - G^{+}_{LR}(\omega)T^{\dagger}]\bsig\}.
\label{eq15}
\end{eqnarray}
\noindent
Again, the charge current is given by Eq.(15) with $\frac {1}{2} \bsig $ replaced by the unit matrix multiplied by $e/\hbar$. 

Similarly, the total spin angular momentum on atomic planes on either side of the cleavage plane, in the non-equilibrium state, is given by
\begin{eqnarray}
<\bS _{n-1}> = -\frac {1}{2} i\hbar \sum_{\sbkp} \int \frac{d\omega}{2\pi}Tr \{G^{+}_{LL}(\omega)\bsig \}
\label{eq16}
\end{eqnarray}
\noindent
\begin{eqnarray}
<\bS _{n}> = -\frac {1}{2} i\hbar \sum_{\sbkp} \int \frac{d\omega}{2\pi}Tr \{G^{+}_{RR}(\omega)\bsig \}.
\label{eq17}
\end{eqnarray}
\noindent
Following Keldysh \cite{keldysh}, \cite{caroli}, we now write
\begin{eqnarray}
G^{+}_{AB}(\omega) = \frac {1}{2}(F_{AB}+G^{a}_{AB}-G^{r}_{AB}),
\label{eq18}
\end{eqnarray}
\noindent
where the suffices $A$ and $B$ are either $R$ or $L$.
$F_{AB}(\omega)$ is the Fourier transform of
\begin{eqnarray}
F_{AB}(t,t') = -i<[c_{A}(t),c^{\dagger}_{B}(t')]_{-}>
\label{eq19}
\end{eqnarray}
\noindent
and $G^{a}$, $G^{r}$ are the usual advanced and retarded Green functions \cite{book}. Note that in \cite{keldysh}and \cite{caroli} the definitions of $G^{a}$ and $G^{r}$ are interchanged and that in the Green function matrix defined by these authors $G^{+}$ and $G^{-}$ should be interchanged.

Charge and spin current, and spin density, are related by Eqs.(15)-(17) to the quantities $G^{a}$, $G^{r}$, and $F_{AB}$. The latter are calculated for the coupled system by starting with decoupled left and right systems, each in equilibrium, and turning on the hopping between planes $L$ and $R$ as a perturbation. Hence, we express $G^{a}$, $G^{r}$, and $F_{AB}$ in terms of retarded surface Green functions $g_{L}\equiv g_{LL}$, $g_{R}\equiv g_{RR}$ for the decoupled equilibrium system. The final result for the spin angular momentum on plane $n$ to the right of the cleavage plane is 
\begin{eqnarray}
<\bS _{n}> = <\bS _{n}>_{1} + <\bS _{n}>_{2},
\label{eq20}
\end{eqnarray}
\noindent
where the two contributions to the spin angular momentum $<\bS _{n}>_{1}$ and $<\bS _{n}>_{2}$ are given by
\begin{eqnarray}
<\bS _{n}>_{1} = -\frac {\hbar}{4\pi} \sum_{\sbkp} \int d\omega \,Im \,Tr \{Ag_{R}\bsig\}[f(\omega-\mu_{L})+f(\omega-\mu_{R})]
\label{eq21}
\end{eqnarray}
\begin{eqnarray}
<\bS _{n}>_{2} = -\frac {\hbar}{2\pi} \sum_{\sbkp} \int d\omega \,Im \,Tr \{(A-\frac{1}{2})Bg_{R}^{\dagger}\bsig\}[f(\omega-\mu_{L})-f(\omega-\mu_{R})].
\label{eq22}
\end{eqnarray}
\noindent
Here, $A=[1-g_{R}T^{\dagger}g_{L}T]^{-1}$, $B=[1-g_{R}^{\dagger}T^{\dagger}g_{L}^{\dagger}T]^{-1}$, and $f(\omega-\mu)$ is the Fermi function with chemical potential $\mu$ and $\mu_{L}-\mu_{R}=eV_{b}$. To obtain $<\bS _{n-1}>$, defined by Eq.(16), we must interchange $L$ and $R$, and $T$ and $T^{\dagger}$, everywhere in Eq.(20)-(22). In the linear-response case of small bias which we are considering, the Fermi functions in Eq.(22) are expanded to first order in $V_{b}$. Hence the energy integral is avoided, being equivalent to multiplying the integrand by $eV_{b}$ and evaluating it at the common zero-bias chemical potential $\mu_{0}$.

As shown in Fig.1, the magnetization $\bP$ of the polarizing ferromagnet is assumed to be fixed in the $(z,x)$ plane and makes an angle $\theta$ with the $z$ axis, which is the direction of the uniaxial anisotropy field in the switching magnet. When a bias is applied, spin-polarized current flows through the switching magnet and exerts a torque on its magnetization. This torque is in competition with the torque due to the anisotropy field and causes the spin  $<\bS _{n}>$ in a given atomic plane $n$ to deviate from the anisotropy axis. In the steady state $<\bS _{n}>$ settles in a definite direction specified by the angles $\alpha_{n}$, $\phi_{n}$ shown in Fig.1. To determine these angles, we assume the exchange field $\bDel_{n}$ in plane $n$ is in the direction $(\alpha_{n},\phi_{n})$ and apply the self-consistency condition
\begin{eqnarray}
\bDel_{n}\times <\bS _{n}> = 0.
\label{eq23}
\end{eqnarray}
\noindent
This condition guarantees that the local magnetization is in the direction of the exchange field, as it should be in the unrestricted Hartree-Fock approximation mentioned in Sec.1. As with anisotropy fields, the exchange field $\Delta _{n}$ is defined as an angular frequency so that $\hbar \Delta_{n}$ is the energy to reverse the spin on plane $n$. More precisely, the spin-dependent part of the on-site energy on plane $n$ is given by $-(1/2)\hbar (\bH_{A} + \bDel_{n}).\bsig$. We assume that $\mid \bDel_{n} \mid$ always takes its bulk value.

Following the method outlined for obtaining Eq.(20), similar expressions in terms of retarded surface Green functions may be obtained for the spin currents $<\bj _{n-1}>$ and $<\bj _{n}>$. Writing again 
$<\bj _{n}>=<\bj _{n}>_{1}+<\bj _{n}>_{2}$, we obtain
\begin{eqnarray}
<\bj _{n-1}>_{1} = \frac{1}{4\pi}\sum_{\sbkp} \int d\omega \, Re Tr\{(B-A)\bsig \}[f(\omega-\mu_{L})+f(\omega-\mu_{R})]
\label{eq24}
\end{eqnarray}
\noindent
\begin{eqnarray}
<\bj _{n-1}>_{2} = \frac{1}{2\pi}\sum_{\sbkp} \int d\omega \, Re Tr\{[g_{L}TABg^{\dagger}_{R}T^{\dagger}-AB+\frac{1}{2}(A+B)]\bsig \}[f(\omega-\mu_{L})-f(\omega-\mu_{R})].
\label{eq25}
\end{eqnarray}
\noindent
By considering the changes in $g_{L}$, $g_{R}$ when the cleavage plane is moved one atomic plane to the right, it is straightforward to show that
\begin{eqnarray}
<\bj _{n-1}> - <\bj _{n}> = (\bH_{A} + \bDel_{n})\times <\bS _{n}>.
\label{eq26}
\end{eqnarray}
\noindent
This equation holds for a steady state with arbitrary exchange fields $\bDel_{n}$ which do not necessarily satisfy the self-consistency condition (23). When the self-consistency condition (23) is satisfied, we recover
the steady-state result 
\begin{eqnarray}
<\bj _{n-1}> - <\bj _{n}> = \bH_{A} \times <\bS _{n}>
\label{eq27}
\end{eqnarray}
\noindent
which was derived earlier [Eq.(12)] purely from considerations of the spin-rotational symmetry of the electron-electron interactions. This verifies the consistency of the Keldysh formalism combined with the unrestricted Hartree-Fock approximation.

It follows from Eqs.(26) and (27) that all components of spin current are conserved within the spacer and  lead, where $\bH_{A}=0, \bDel_{n}=0$, with or without self-consistency. Furthermore, it follows from Eq.(27) that in the self-consistent steady state the component of spin current in the direction of the anisotropy field $\bH_{A}$ is conserved throughout the system, as of course is the charge current. If Eq.(27) is summed over all planes in the switching magnet, we obtain
\begin{eqnarray}
<\bj _{spacer}> - <\bj _{lead}> = \bH_{A} \times <\bS _{tot}>,
\label{eq28}
\end{eqnarray}
\noindent
where $<\bj _{spacer}>$, $<\bj _{lead}>$ are the spin currents in the spacer and lead, respectively, and $<\bS _{tot}>$ is the total spin angular momentum of the switching magnet. This shows how the total spin transfer torque acting on the switching magnet is balanced by the torque exerted by the anisotropy field on the total moment.

We have separated in Eq.(20) the spin angular momentum $<\bS _{n}>$ into two parts $<\bS _{n}>_{1}$ and $<\bS _{n}>_{2}$. It is clear that $<\bS _{n}>_{2}$ is proportional to the applied bias $V_{b}$ to the first order and for zero bias ($\mu_{L}=\mu_{R}$) only $<\bS _{n}>_{1}$ remains. The spin transfer torque $<\bj _{n}> - <\bj _{n-1}>$ similarly splits into two parts (Eqs.(24),(25))
in such a way that Eq.(26) holds for each component separately:
\begin{eqnarray}
<\bj _{n-1}>_{i} - <\bj _{n}>_{i} = (\bH_{A}+\bDel_{n}) \times <\bS _{n}>_{i}, \,\,i=1,2. 
\label{eq29}
\end{eqnarray}
\noindent
Only the first part $<\bj _{n-1}>_{1} - <\bj _{n}>_{1}$ is nonzero at zero bias. It corresponds to spin currents which mediate exchange coupling, either between the two magnets across the spacer or between atomic planes in the switching magnet. Consequently, at zero bias the spin current in the lead is zero. It is easy to verify that the expressions for interlayer exchange coupling derived here, using the Keldysh formalism, agree precisely with those obtained earlier by other methods \cite{exchcp}.

The results of this section show the great advantage of the Keldysh formalism, even within the linear response regime. Spin currents at zero bias, corresponding to exchange coupling, transport spin and particle currents, and spin densities are all calculated in a unified way. Relationships between these quantities, such as Eq. (29), are then easily derived. In the standard linear response theory of Kubo zero-bias quantities cannot be calculated and different response functions would have to be introduced for calculating currents and spin density response at finite bias.   

\section{Application to a switching monolayer}

In the general theory outlined in Sec.3 the steady-state spin orientation of each atomic plane $n$ of the switching magnet must be determined self-consistently. In this section we first consider the simplest case of a single orbital on each site and when the switching magnet is a single atomic plane. In this case there is no interplane exchange coupling in the switching magnet to consider and we assume that the spacer is sufficiently thick for the zero-bias exchange coupling between the two ferromagnets to be negligible. For a given bias $V_{b}$, the direction $(\alpha_{0},\phi_{0})$ of the steady-state orientation of the  switching magnet moment $<\bS>$
is determined self-consistently from Eq.(23) with the cleavage plane immediately to the left of the switching plane so that $<\bS_{n}>=<\bS>$.
It is convenient to determine $(\alpha_{0},\phi_{0})$ in two steps. The first step locates a 'universal path' on a unit sphere, independent of $V_{b}$, on which the self-consistent solutions for any given $V_{b}$ must lie. In the second step the bias $V_{b}$ required to stabilize the magnetization in a given direction $(\alpha_{0},\phi_{0})$ is determined as a function of $\alpha_{0} $, say. To establish this program, we write Eq.(23) as 
\begin{eqnarray}
(\bH_{A} + \bDel )\times <\bS> = \bH_{A}\times <\bS>,
\label{eq30}
\end{eqnarray}
\noindent
where $\bDel $ is the exchange field of the switching layer in the direction 
$(\alpha_{0},\phi_{0})$. Splitting $<\bS>$ into two parts as in Eq.(20), this becomes
\begin{eqnarray}
(\bH_{A} + \bDel )\times <\bS>_{1} + (\bH_{A} + \bDel )\times <\bS>_{2} = \bH_{A}\times <\bS>.
\label{eq31}
\end{eqnarray}
\noindent
Hence using Eq.(29) we have 
\begin{eqnarray}
<\bj _{spacer}>_{1} - <\bj _{lead}>_{1} + <\bj _{spacer}>_{2} - <\bj _{lead}>_{2} = \bH_{A}\times <\bS>.
\label{eq32}
\end{eqnarray}
\noindent
The first two terms on the left correspond to exchange coupling torque which, as discussed above, is assumed to be negligible compared with the anisotropy torque. This is justified for thick spacers since the interlayer exchange coupling tends to zero as the spacer thickness tends to infinity. 
The last two terms on the left correspond to the spin transfer torque $\bT$, which is proportional to bias $V_{b}$, and the right-hand side of Eq.(32) is $-\bT_{A}$, where  $\bT_{A}$ is the torque exerted by the anisotropy field on the switching magnet. We shall see that, in contrast to the exchange coupling torque, $\bT$ remains finite as spacer thickness tends to infinity. Hence 
\begin{eqnarray}
\bT = <\bj _{spacer}>_{2} - <\bj _{lead}>_{2} = \bH_{A}\times <\bS> = -\bT_{A}
\label{eq33}
\end{eqnarray}
\noindent
and, in particular,
\begin{eqnarray}
\bT .\bH_{A} = 0.
\label{eq34}
\end{eqnarray}
\noindent
The bias $V_{b}$ now cancels and this equation determines the universal path described above. Eq.(33) determines the bias required to stabilize any particular point on this path of possible steady states.

We conclude this section with one example in which, for simplicity, we retain only uniaxial anisotropy, this field being chosen in the z direction. We use a single-orbital tight-binding  model whose lattice is taken to be simple cubic with layering in the (010) direction. The nearest-neighbor hopping parameter $t$ is taken to be the same throughout the system. The on-site energy in the spacer and lead is taken as $V_{sp}$, the zero of energy being at the common Fermi level for zero bias. In this example the on-site energy $V_{sp}$ is also taken for majority spin in the ferromagnets (perfect matching in the majority-spin channel). The on-site energy for minority spin in the ferromagnets is taken as $V_{sp} + \hbar \Delta$, where $\hbar \Delta$ is the exchange splitting. The matching of spacer and majority spin bands is similar to the situation in Co/Cu. We take $V_{sp}=2.3, \hbar \Delta =0.7$ in units of $2t$. Furthermore, we take the uniaxial field parameter $H_{u0}=1.86\times 10^{10}sec^{-1}$ which corresponds to a field of $0.106T$. We also take a general value $\theta =2$ radians of the angle between the polarizing magnet moment and the direction of the uniaxial anisotropy axis of the switching magnet. To determine the torque $\bT$ which appears in Eq.(34) for the universal path, we need to calculate the Green functions $g_{L}, g_{R}$ which are required in Eq.(25). This is done by standard adlayering methods described previously \cite{prb97}. At this stage, the anisotropy field is included in the calculation of all the Green functions. Fig.2 shows the calculated universal path of $\phi $ versus $\alpha$ for the
\begin{figure}[here]
\centerline{\epsfig{figure=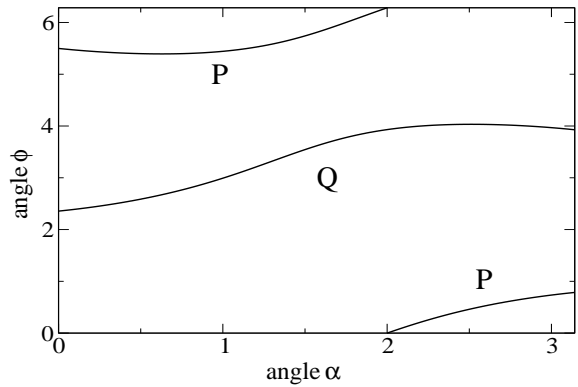,width=3in,height=2in,angle=0}}
\caption{\footnotesize Universal path of $\phi$ versus $\alpha$.} 
\end{figure}
\noindent 
specified parameters. The bias $V_{b}$ required to yield a steady-state magnetization at a given point of this path is plotted as a function of $\alpha$ in Fig.3, where we have assumed the band width $12t=6$eV. Positive bias corresponds to a drop in voltage between the polarizing magnet and the lead. The correspondence between the two curves in Fig.2 and Fig.3 is indicated by letters P and Q. The discussion of stability of these steady states and the interpretation of Fig.3 is postponed to section 7.
\begin{figure}[here]
\centerline{\epsfig{figure=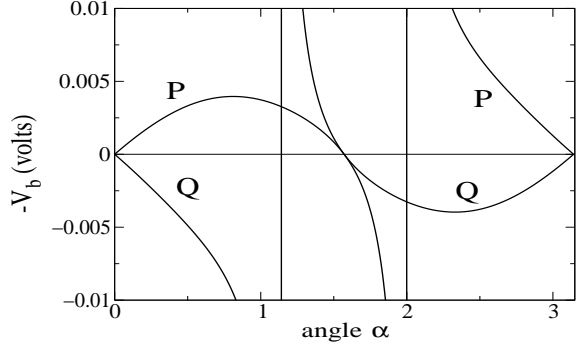,width=3in,height=2in,angle=0}}
\caption{\footnotesize Bias $V_{b}$ required to stabilize the switching magnet moment  at an angle $\alpha$ on the universal path.} 
\end{figure}
\noindent  
The method of calculating steady states used in this section becomes more complicated when the switching magnet contains several atomic planes since the moments $<\bS_{n}>$ of all planes must be determined self-consistently. This entails the inclusion of the exchange stiffness between atomic planes of the switching magnet which is contained in Eq.(29) with $i=1$. To address this problem, we introduce in the next section the simplifications required to derive from first principles the convenient 'standard model' used by previous authors \cite{slon}, \cite{sun}, \cite{stiles}.

\section{The standard model}

In the Keldysh method used above it is essential to include the anisotropy field $\bH_{A}$ in the Hamiltonian to obtain a non-trivial steady state. In the absence of $\bH_{A}$ it follows from Eq.(27) that in the steady state all components of spin current are conserved everywhere so that there are no spin-transfer torques. Hence the only steady state is the trivial one in which the switching magnet is aligned parallel or antiparallel to the polarizing magnet. Previous authors \cite{slon}, \cite{stiles} did not consider a steady state but calculated spin-transfer torque as a one-electron problem with the exchange fields of the polarizing and switching magnets at a fixed assumed angle. In a second independent step, these authors \cite{slon}, \cite{sun} balance the spin-transfer torque against the torque due to anisotropy field in the context of a Landau-Lifshitz equation. This approach is what we call the standard model (SM). In this section we show how to arrive at the SM by making some simplifying approximations in our self-consistent approach. 

We begin with the monolayer switching magnet of Sec.4. In Eq.(33) the spin-transfer torque $\bT$ is calculated in the presence of $\bH_{A}$ and the spin $<\bS>$ is the self-consistent moment. To obtain the SM we must neglect $\bH_{A}$ in the calculation of $\bT$ and replace $<\bS>$ by its nonself-consistent value in the direction of the assumed exchange field of the switching magnet. These approximations are both reasonable provided the exchange field is much stronger than the anisotropy field, which is satisfied for a ferromagnet such as Co. This follows since the Green functions which determine both $\bT$ and $<\bS>$ depend on the total field $\bDel + \bH_{A}$. Thus in the SM the spin-transfer torque is calculated as a function of the angle between the magnetizations without solving the  self-consistency problem. Furthermore, equating it to the anisotropy torque as in Eq.(33) is equivalent to calculating a steady state
of the Landau-Lifshitz equation. The justification of the SM for a switching magnet with more than one atomic plane is more subtle.

The self-consistency condition (23) must be satisfied for each plane in the switching magnet. It may be written [c.f. Eq.(30) for the monolayer]
\begin{eqnarray}
(\bH_{A} + \bDel_{n})\times <\bS_{n}> = \bH_{A}\times <\bS_{n}> 
\label{eq35}
\end{eqnarray}
\noindent
and, using again $<\bS _{n}> = <\bS _{n}>_{1} + <\bS _{n}>_{2}$ and Eq.(29), we obtain
\begin{eqnarray}
<\bj _{n-1}>_{1} - <\bj _{n}>_{1} + <\bj _{n-1}>_{2} - <\bj _{n}>_{2}= \bH_{A}\times <\bS_{n}>. 
\label{eq36}
\end{eqnarray}
\noindent
The first two terms contain the interlayer exchange coupling, which is neglected as in the monolayer case, and interplane exchange coupling within the switching magnet. To clarify the argument, we write notionally this last contribution in terms of local exchange stiffness $D_{n}$ between atomic planes $n-1$ and $n$. Hence from Eq.(36)
\begin{eqnarray}
 <\bj _{n-1}>_{2} - <\bj _{n}>_{2}= \bH_{A}\times <\bS_{n}> + D_{n-1}(\bS_{n-1}\times\bS_{n}) - D_{n}(\bS_{n}\times\bS_{n+1}). 
\label{e37}
\end{eqnarray}
\noindent
On summing over all planes $n$ in the switching magnet the internal exchange coupling torques cancel and we have
\begin{eqnarray}
 <\bj _{spacer}>_{2} - <\bj _{lead}>_{2}= \bH_{A}\times \sum_{n}<\bS_{n}>. 
\label{eq38}
\end{eqnarray}
\noindent
In the fully self-consistent solution of Eq.(35) the exchange field $\bDel_{n}$ is parallel to the local moment $<\bS_{n}>$ but the $\bDel_{n}$ are not collinear. To proceed to the SM we must assume that all the $\bDel_{n}$ used to calculate the spin-transfer torque on the left of Eq.(38) are equal, to $\bDel$, say. Furthermore, we assume, as in the case of the monolayer, that we can neglect $\bH_{A}$ in the calculation of the spin-transfer torque and that all $<\bS_{n}>$ are in the direction $\bDel$ with magnitude equal to the ground state moment. In making the approximation $\bDel_{n}\approx \bDel$ we have failed to satisfy Eqs.(37) individually as is required for full self-consistency.

To show that this is not serious for a ferromagnet such as Co, consider the following argument. If we use the uniform value of $\bDel$ determined from Eq.(38), as described, to calculate the spin-transfer torque in Eq.(37) and assume all $<\bS_{n}>$  are in the direction of $\bDel$, the last two terms of Eq.(37) are zero and the equations are far from satisfied. However, since the exchange stiffness constants $D_{n}$ of a ferromagnet such as Co are large one need only introduce small deviations of $\bDel_{n}$ from the uniform $\bDel$, and consequently small deviations of  $<\bS_{n}>$ from uniformity, to satisfy the self-consistent equations (37). This is true because the spin-transfer and anisotropy torques are insensitive to these small deviations. The ability of the SM to simulate the fully self-consistent solution accurately has been verified numerically for a switching magnet with two atomic planes using the single-orbital model of Sec.4.

\section{Two components of the spin-transfer torque in the standard model}

In the calculation of the spin-transfer torque $\bT$ within the standard model the anisotropy field is neglected so that $\bT$ depends only on the angle $\psi$ between the magnetization $\bP$ of the polarizing magnet and the assumed exchange field $\bDel$ of the switching magnet. As in Fig.1, the magnetization $\bP$ of the polarizing magnet is in the (z,x) plane, making an angle $\theta$ with the z axis, and, for convenience, we choose 
the exchange field of the switching magnet to be in the z direction so that $\psi = \theta $. The torque $\bT$ in the SM is given by
\begin{eqnarray}
\bT = <\bj _{spacer}>_{2} - <\bj _{lead}>_{2} 
\label{eq39}
\end{eqnarray}
\noindent
where the right-hand side is related to the total bias-induced spin $\bS_{2}=\sum_{n}<\bS_{n}>_{2}$ by Eq.(29), summed over $n$, with $\bH_{A}$ neglected and $\bDel _{n}=\bDel$. It follows that 
\begin{eqnarray}
\bT = \bDel \times  \sum_{n}<\bS_{n}>_{2} = \bDel \times \bS_{2}.
\label{eq40}
\end{eqnarray}
\noindent
Here, $<\bS_{n}>_{2}$ is given by Eq.(22) with $\bH_{A}$ neglected in the Green functions $g_{L}$, $g_{R}$. The three components of $\bS_{2} = (S_{2x},S_{2y},S_{2z})$ are related to the three Pauli matrices $\sigma _{x}, \sigma _{y}, \sigma _{z}$ in Eq.(22). Clearly, from Eq.(40) the z component of torque is zero so that we can write
\begin{eqnarray}
\bT = (T_{x},T_{y},0).
\label{eq41}
\end{eqnarray}
\noindent
The 'in-plane' component $\Tpar =T_{x}$, where 'in-plane' refers to the (z,x) plane containing $\bP$ and $\bDel$, is given by $\Tpar=-\Delta S_{2y}$ and the 'out-of-plane' component $\Tper =T_{y}$ is given by  $\Tper = \Delta S_{2x}$. The quantities $S_{2x}$ and $S_{2y}$ represent small deviations of the switching magnet moment from the direction of its exchange field. These spin components are referred to by previous authors \cite{fert}, \cite{heide} as 'spin accumulation'. In the self-consistent steady-state treatment of Sec.4 such deviations do not occur because the exchange field is always in the direction of the local moment. In our view, time-independent spin accumulation $S_{2x}$, $S_{2y}$ in the ferromagnet is a non-physical concept which, however, we may define formally as the ratio of torques $\Tpar$, $\Tper$, to the exchange field $\Delta$. It is remarkable that, as shown in Sec.5, the SM in which this concept arises provides a convenient and frequently accurate method for calculating spin-transfer torque. Time-dependent spin accumulation in the ferromagnet in a non-steady state could be a valid concept. However, these time-dependent spins would produce time-dependent exchange fields which would excite the whole spin system. This would require a many-body treatment going beyond the unrestricted Hartree-Fock approximation which is adequate for the steady state. Spin accumulation $<S_{x}>$, $<S_{y}>$, $<S_{z}>$ proportional to the bias exists in the spacer even in the steady state and has real physical significance. Calculation of this effect will be published in a succeeding paper.

The spin transfer torque can be calculated either directly from Eq.(39) or from Eq.(40). However, the latter would require calculating $<\bS_{n}>_{2}$ for each atomic plane of the switching magnet so that the direct method is obviously preferable. 

We begin with an exactly solvable one-band model which we can connect with previous work \cite{slon}. This model is related to the one described at the end of Sec.4, where the switching magnet is a single atomic plane and there is perfect matching between the spacer band and the majority spin bands in both ferromagnets. To obtain analytical results in this first example, we also assume that the exchange splitting $\Delta \rightarrow \infty$ both in the polarizing and switching magnets. In fact, once the bottom of the minority-spin band   is well above the Fermi level, the results are rather insensitive to the magnitude of $\Delta$. Such a system is sometimes referred to as a half-metallic ferromagnet and is the first case considered  by Slonczewski in his original paper \cite{slon}. In the limit $\Delta \rightarrow \infty$ the SM model is exact since the moment of the switching magnet cannot deviate from the exchange field and the self-consistency condition (23) is automatically satisfied. We, therefore, calculate the spin-transfer torque in the absence of anisotropy field. Clearly, owing to the infinite exchange splitting, the only spin current in the lead corresponds to the z component of spin  and the z-spin current is equal to the charge current (multiplied by $\hbar /2e)$. It turns out in this model that the y-spin current in the spacer, which is equal to the torque $\Tper$ since the corresponding current in the lead is zero, vanishes. Thus only Slonczewski torque $\Tpar$ survives and is given by
\begin{eqnarray}
\bTpar = -\be_{x} \frac{eV}{8\pi} \sum _{\sbkp}\frac{t^{2}(g_{0}-g_{0}^{*})^2 \sin \psi}{\mid 1-t^{2}g_{0}(a+b\cos\psi)\mid^{2}}.
\label{eq42}
\end{eqnarray}
\noindent
Here, $g_{0}=g_{0}(\bkp,0)$ is the majority spin surface Green function for the semi-infinite ferromagnet, or equivalently for the semi-infinite ferromagnet with an overlayer of the matching spacer. The Green function $g_{0}$ is evaluated at energy $\omega =0$, the common Fermi level of the unbiased system. Also, $t$ is the hopping parameter introduced in Sec.4 and $a$ and $b$ are given by
\begin{eqnarray}
a=\frac{1}{2}(g_{0}+A_{N})\ \ ;\ \  b=\frac{1}{2}(g_{0}-A_{N}),
\label{eq43}
\end{eqnarray}
\noindent
where $A_{N}=\sin Nk_{\perp}d/[t\sin(N+1)k_{\perp}d]$ with
\begin{eqnarray}
k_{\perp}d = \cos ^{-1}[(V^{\uparrow}_{s}+2t(\cos k_{x}d+\cos k_{y}d))/2t]
\label{eq44}
\end{eqnarray}
\noindent
and $\bkp=(k_{x},k_{y},0)$; $d$ is the interatomic distance, N is the number of atomic planes in the spacer, and $V^{\uparrow}_{s}$ is the on-site potential in the majority-spin band of the switching magnet. The corresponding expression for the charge current is very similar and we find that
\begin{eqnarray}
 \Tpar = \frac{\hbar}{2\mid e \mid}\tan(\psi /2) \times (charge \; current).
 \label{eq45}
\end{eqnarray}
\noindent
This is precisely the Slonczewski result for the analogous parabolic band model.
It should be noted that the torque $\Tpar $ goes to zero for $\psi \rightarrow \pi$ since the charge current for a halfmetallic magnet contains a factor $1+\cos\psi$.

The interesting result that $\Tper =0$ for this model may be traced to an effective reflection symmetry of the system about a plane at the center of the spacer. Although the present system appears asymmetric the infinite exchange splitting makes it equivalent to a symmetric system with semi-infinite switching magnet. More generally, we find, certainly for a one-band model with arbitrary parameters, that the y-spin component of the spin current in the spacer always vanishes for a system with reflection symmetry. In general, however, the y-spin current in the lead is non-zero so that $\Tper \neq 0$ even for a symmetric system. The result $\Tper =0$ for the above model is, therefore, a very special one due to the artefact of a very large exchange splitting in the ferromagnets. 

In the second set of examples we consider several cases, within the one-band model, where the exchange splittings in the ferromagnets are finite. Simple formulas, such as Eq. (42), for the torques are no longer available and they must be calculated numerically. In all the examples the calculated torques are per surface atom.  In all cases, we retain the geometry of the first example.  Table I lists the parameters for all the cases considered.

\begin{table}
\caption{Parameters for one-band models. $V_{p}^{\uparrow}$, $V_{p}^{\downarrow}$ are on-site potentials for majority and minority spin in the polarizing magnet; $V_{s}^{\uparrow}$, $V_{s}^{\downarrow}$ are on-site potentials for majority and minority spin in the switching magnet; $V_{sp}$ is the on-site potential in the spacer and the lead; $N$, $M$ are the numbers of atomic planes in the spacer and switching magnet, respectively. }
\vspace{0.2cm}
\begin{tabular}{|c|c|c|c|c|c|c|c|}
\hspace{0.3cm} $case$ \hspace{0.3cm} & \hspace{0.3cm} $V_{p}^{\uparrow}$ \hspace{0.3cm} & \hspace{0.3cm}$V_{p}^{\downarrow}$\hspace{0.3cm} & \hspace{0.3cm}$V_{sp}$\hspace{0.3cm} & \hspace{0.3cm} $V_{s}^{\uparrow}$\hspace{0.3cm} & \hspace{0.3cm}$V_{s}^{\downarrow}$\hspace{0.3cm}
 & \hspace{0.3cm} $N$\hspace{0.3cm} & \hspace{0.3cm}$M$ \hspace{0.3cm} \\ 
\hline \hline
(a) & 2.3 & 3.0 & 2.3 & 2.3 & 5.0 & 20 & 1 \\
(b) & 2.3 & 3.0 & 2.3 & 2.3 & 3.0 & 20 & 1 \\
(c) & 2.3 & 3.0 & 2.3 & 2.8 & 3.0 & 20 & 1 \\
(d) & 2.1 & 3.0 & 2.8 & 2.1 & 5.0 & 20 & 1 \\
(e) & 2.3 & 3.0 & 2.3 & 2.8 & 3.0 & 20 & 1-10\\ 
\end{tabular}
\label{tab1}
\end{table}   
\noindent
All potentials in Table 1 are in units of $2t$ and the Fermi energy $\mu_{0}=0$.

Figure 4 (a)  shows the calculated torques $\Tpar$ per surface atom (in units of $eV_{b}$) 
as a function of the angle $\psi$ for the models with parameter sets (a)-(d) of Table 1. 
\begin{figure}[here]
\centerline{\epsfig{figure=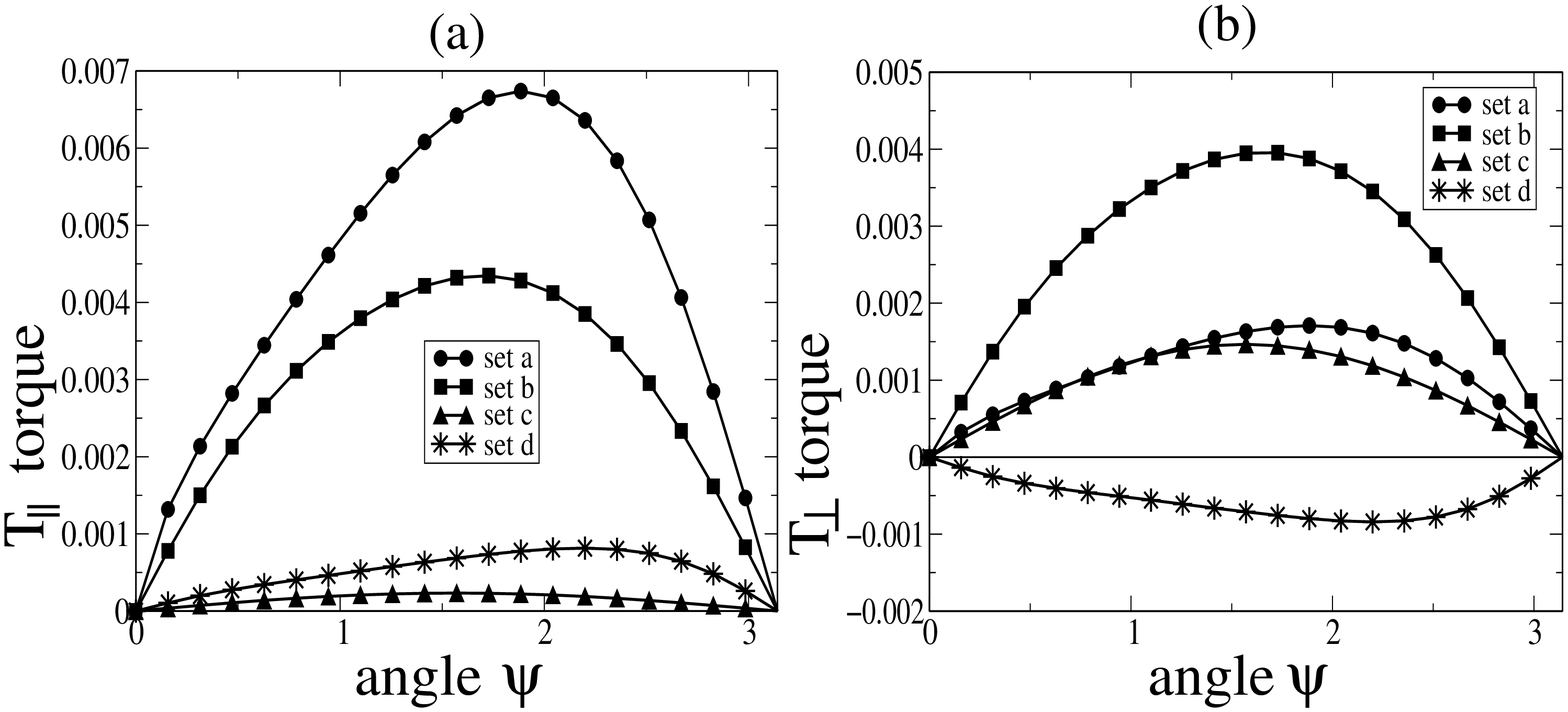,width=4.5in,height=2in,angle=0}}
\vspace{0.5cm}
\caption{\footnotesize Dependence of the spin-transfer torques $\Tpar$ (a) and $\Tper$ (b) on the angle $\psi$ for the model with parameter sets (a)-(d) of Table 1. The torques are in units of $eV_{b}$.} 
\end{figure}
\noindent
In case (a) $\Tpar$ dominates, as expected from the large exchange splitting $V_{s}^{\uparrow} - V_{s}^{\downarrow}$ in the switching magnet which approaches the infinite exchange splitting of our first exactly solvable model. The angular dependence of both torques is clearly dominated by a $\sin \psi $ factor (c.f. Eq.(42)) although some distortions are apparent. In case (b)
$\Tpar$ and $\Tper$ are of almost equal strength. This is case where the two ferromagnets are of the same material and the bottom of the minority spin band is exactly at the Fermi level. This simulates well the situation in Co/Cu and we shall see presently that in realistic calculations for this system the torques $\Tpar$ and $\Tper$ are again similar in magnitude. The parameters of case (b) were used previously in Sec.4 as an example of a fully self-consistent calculation of steady states. In case (c)  $\Tper$ is larger than $\Tpar$. 
It is interesting that this occurs for smaller exchange splitting in the switching ferromagnet. In cases (b) and (c) the angular dependence of the torques is hardly distorted from the $\sin \psi $ form. In cases (a), (b), (c)
the two torques $\Tpar$, $\Tper$ have the same sign. In case (d) they have opposite sign and almost equal magnitude. In examples (a)-(d)
the switching magnet consists of one atomic plane. In case (e) shown in Fig. 5 we use the same parameters as in case (c) but the number of atomic planes in the switching magnet varies between 1 and 10 and the angle $\psi=\pi/2$.
\begin{figure}[here]
\centerline{\epsfig{figure=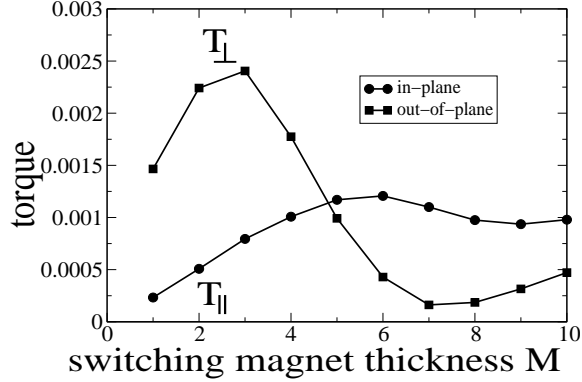,width=3in,height=2in,angle=0}}
\vspace{0.5cm}
\caption{\footnotesize Dependence of the spin-transfer torque $\Tpar$ and $\Tper$ on the thickness of the switching magnet $M$ for
 $\psi =\pi/2$ and the parameter set (c) of Table 1. The torques are in units of $eV_{b}$.} 
\end{figure}
\noindent
It can be seen from Fig.5 that, contrary to popular belief, the out-of-plane torque $\Tper$ dominates over $\Tpar$ for small thicknesses of the switching magnet and remains 50\% of $\Tpar$ at $M=10$ at. planes. For all thicknesses of the switching magnet, $\Tpar$ and $\Tper$ have the same sign. We have already  seen  in case (d) that this is not always the case.

So far we have kept the number of atomic planes in the spacer at 20 but we must now highlight an important and surprising result concerning the dependence of torque on $N$. In Fig.6 we show the torques for the parameter set (b) and $\psi = \pi/2$ plotted as functions of $N$. 
\begin{figure}[here]
\centerline{\epsfig{figure=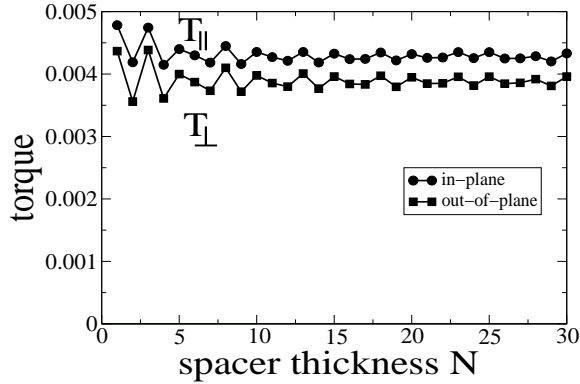,width=3in,height=2in,angle=0}}
\vspace{0.5cm}
\caption{\footnotesize Dependence of the spin-transfer torque $\Tpar$ and $\Tper$ on the thickness of the spacer $N$ for
 $\psi =\pi/2$ and the parameter set (b) of Table 1. The torques are in units of $eV_{b}$.} 
\end{figure}
\noindent
It is clear that they both oscillate but tend to constant values as $N\rightarrow \infty$ in our ballistic limit. In zero bias all spins of the system lie in the (z,x) plane and it is, therefore, not surprising that when charge current flows in nonzero bias there are finite z-spin and x-spin currents for arbitrary spacer thickness. The constant value of y-spin current as $N\rightarrow \infty$, this being associated with $\Tper$ , may seem more surprising since there is no y-spin density in zero bias. Of course, for $\psi \neq 0$, there is a y-spin current even in zero bias corresponding to interlayer exchange coupling, but this effect  is not associated with charge transport and decays as $1/N^{2}$ with increasing spacer thickness. The relations $\Tpar = -\Delta S_{2y}$, 
$\Tper = \Delta S_{2x}$ derived in the beginning of this section within the standard model show that bias-induced in-plane spin density is related to out-of-plane torque and vice versa. It is, therefore, inevitable that both torques will exist.

As a final example of these spin-transfer torque calculations we consider a fully realistic multi-orbital model of fcc Co/Cu(111) with tight-binding parameters fitted to the results of first-principles band structure calculations, as described previously \cite{prb97}. Referring to Fig.1, the polarizing magnet is a semi-infinite slab of Co, the spacer is 20 atomic planes of Cu, the switching magnet contains  $M$ atomic planes of Co with $M$=1 and 2, and the lead is semi-infinite Cu. Figure 7(a),(b) shows the angular dependences
\begin{figure}[here]
\centerline{\epsfig{figure=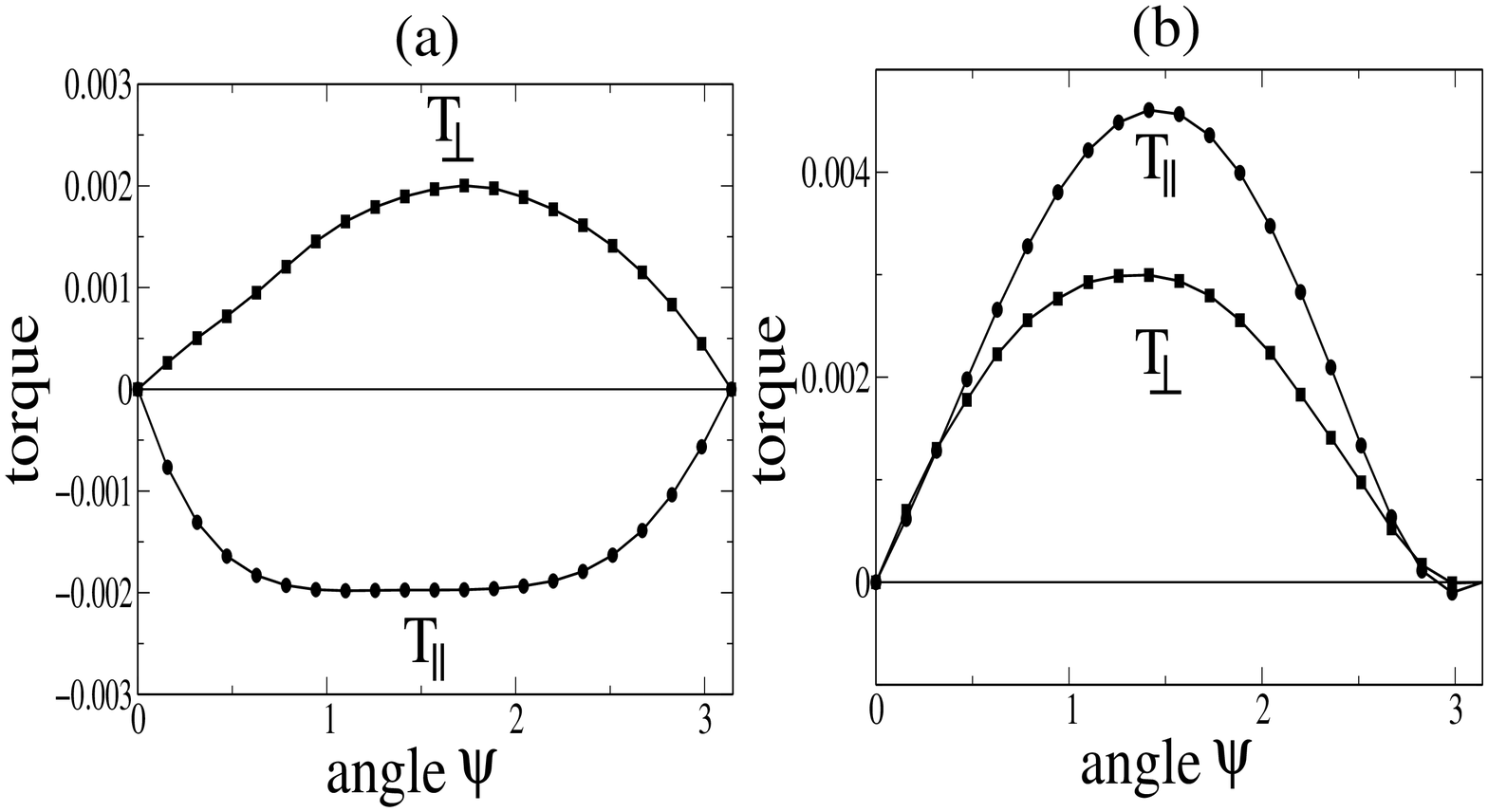,width=5in,height=2in,angle=0}}
\vspace{0.5cm}
\caption{\footnotesize Dependence of the spin-transfer torque $\Tpar$ and $\Tper$ for Co/Cu/Co(111) on the angle $\psi$. The torques per surface atom are in units of $eV_{b}$. Figure (a) is for M=1, and (b) for M=2 monolayers of Co in the switching magnet.} 
\end{figure}
\noindent
of $\Tpar$, $\Tper$ for the cases $M=1$ and $M=2$, respectively. For the monolayer switching magnet, the torques $\Tper$  and  $\Tpar$ are equal in magnitude and they have the opposite sign. However, for $M=2$, the torques have the same sign and $\Tper$ is somewhat smaller than $\Tpar$.  A negative sign of the ratio of the two  torque components has important and unexpected consequences for hysteresis loops as discussed in the next section. Finally, we show in Fig.8 the dependence of $\Tper$  and  $\Tpar$ on the thickness of the
\begin{figure}[here]
\centerline{\epsfig{figure=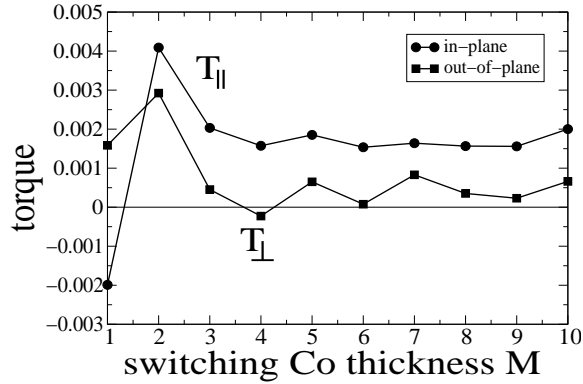,width=3in,height=2in,angle=0}}
\vspace{0.5cm}
\caption{\footnotesize Dependence of the spin-transfer torque $\Tpar$ and $\Tper$ for Co/Cu/Co(111) on the thickness of the switching magnet $M$ for
 $\psi =\pi/3$. The torques are in units of $eV_{b}$.} 
\end{figure}
\noindent
Co switching magnet. It can be seen that the out-of-plane torque $\Tper$ becomes smaller than $\Tpar$ for thicker switching magnets. This is the expected behavior since our polarizing magnet is semi-infinite Co, so that as the switching  Co magnet becomes thicker we approach the limit of a symmetric junction for which the y-component of the spin current vanishes and the corresponding component in the lead is usually small. However, $\Tper$ is by no means negligible (27\% of $\Tpar$) even for a typical experimental thickness of the switching Co layer of ten atomic planes. It is also interesting that beyond the monolayer thickness, the ratio of the two torques is positive with the exception of $M=4$.

\section{Stability of steady states and hysteresis loops}

In Sec.4 we calculated the steady-state orientation of the magnetization of the switching magnet fully self-consistently as a function of bias $V_{b}$ (see Fig.3). These results allow us to calculate the continuous portions of the hysteresis loops of resistance versus bias, but to determine where jumps occur we must investigate the stability of the steady states. This cannot be done within the standard Keldysh formalism since the dynamics of the system, including damping, lies  outside its scope. We, therefore, map the microscopic problem onto a phenomenological time-dependent Landau-Lifshitz (LL) equation with Gilbert damping. An approach based on LL equation has been used previously \cite{slon}, \cite{sun} but our treatment differs in several important ways. Firstly, we use as an input microscopically determined torques $\Tpar$, $\Tper$. In previous work only $\Tpar$ was considered and its magnitude was an adjustable parameter. Secondly, the importance of the steady state concept in the context of the LL equation has not previously been fully recognized. Finally, the consequences of easy plane anisotropy have hardly been explored; we find that, in fact, it can lead to completely new switching scenarios when $\Tpar$ and $\Tper$ have the opposite sign. Even in the absence of easy plane anisotropy, $\Tpar$ and $\Tper$ with the opposite sign may lead to qualitatively different types of switching. 

In Sec.6 the unit vector in the direction of the switching magnet moment 
${\bf m}$ was always taken in the $z$ direction but to discuss the LL equation we must consider ${\bf m}$ in a general direction. The total spin-transfer torque ${\bf T}$ may be written quite generally as the sum of the two components in the directions of the vectors ${\bf m\times p}$ and ${\bf m\times (p\times m)}$, where ${\bf p}$ is  a unit vector in the direction of the magnetization of the polarizing magnet. Thus
\begin{eqnarray}
\bT = \bTper + \bTpar,
\label{eq46}
\end{eqnarray}
\noindent
where
\begin{eqnarray}
\bTper = g_{\perp}\; (\bm \times \bp)
\label{eq47}
\end{eqnarray}
\noindent
\begin{eqnarray}
\bTpar = g_{\parallel}\; \bm \times (\bp \times \bm).
\label{eq48}
\end{eqnarray}
\noindent
The modulus of both vector products in Eqs.(47), (48)   is equal to $\sin \psi$, where $\psi \;(0\le\psi\le\pi)$ is the angle between $\bp$ and $\bm$. Since in the SM $\bTper$ and $\bTpar$ depend only on the angle $\psi$, it follows that the coefficients $g_{\perp}$, $g_{\parallel}$ are functions only of $\psi$. To determine $g_{\perp}(\psi)$, $g_{\parallel}(\psi)$, we return to the geometry of Sec.6 with $\bp=(\sin \psi,0,\cos \psi)$ and $\bm=(0,0,1)$. It follows that
\begin{eqnarray}
\bTpar = g_{\parallel}(\psi)(\sin \psi,0,0)
\label{eq49}
\end{eqnarray}
\noindent
\begin{eqnarray}
\bTper = g_{\perp}(\psi)(0,\sin \psi,0).
\label{eq50}
\end{eqnarray}
\noindent
Thus the scalar quantities calculated in Sec.6 are  $\Tpar = g_{\parallel}\sin \psi$, $\Tper = g_{\perp}\sin \psi$. Hence the magnitudes and signs of $g_{\parallel}(\psi)$, $g_{\perp}(\psi)$ are determined. It is seen in Fig. 4 that the $\sin \psi$ factor accounts for most of the angular dependence of $\Tpar$, $\Tper$ so that to a good approximation $g_{\parallel}$, $g_{\perp}$ are constants, proportional to bias. 

The LL equation takes the form 
\begin{eqnarray}
\frac{d\bm}{dt} + \gamma \bm \times \frac{d\bm}{dt} = \bG
\label{eq51}
\end{eqnarray}
\noindent
with the reduced total torque $\bG$ given by
\begin{eqnarray}
\bG = (-\bH_{A} \times <\bS_{tot}> + \bTper + \bTpar)/\mid <\bS_{tot}> \mid,
\label{eq52}
\end{eqnarray}
\noindent
where $<\bS_{tot}>$ is the total spin angular momentum of the switching magnet and $\gamma$ is the Gilbert damping parameter. Following Sun \cite{sun}, Eq.(51) may be written more conveniently as
\begin{eqnarray}
(1+\gamma ^{2})\frac{d\bm}{dt}  = \bG - \gamma \bm \times \bG.
\label{eq53}
\end{eqnarray}
\noindent
We first consider steady-state solutions of this equation but shall return to the full time-dependent equation when discussing stability of these states.
In the steady state Eq(51) reduces to 
\begin{eqnarray}
-\bH_{A} \times <\bS_{tot}> + \bTper + \bTpar = 0
\label{eq54}
\end{eqnarray}
\noindent
which is equivalent to Eq.(33).

The magnetization unit vector $\bp$ of the polarizing magnet is given by $(\sin \theta ,0,\cos \theta )$ and in the phenomenological treatment, based on the SM, the magnetization of the switching magnet is uniform in the direction $\bm =(\sin \alpha \cos \phi , \sin \alpha \sin \phi , \cos \alpha )$ (see Fig.1). The procedure for finding steady states is exactly analogous to that described in Sec.4 for the microscopic approach. Thus the universal path on the unit sphere consisting of points ($\alpha , \phi $) which correspond to possible steady states, independent of bias, is again given by Eq.(34). For given $\theta $ the torque $\bT = \bTper + \bTpar $ is now defined explicitly as a function of $\alpha $ and $\phi $ by Eqs.(47), (48), the bias factor in the constants $g_{\perp}$, $g_{\parallel}$ cancelling as before. Similarly, using Eq.(54), we can plot $\alpha $ against bias $V_{b}$ for the actual steady state in the given bias as in Fig.3.

We now return to Eq.(51) to discuss stability of the steady states. The torque $\bG$ is given by
\begin{eqnarray}
\bG = H_{u0}\{(\bm.\be_{z})\bm \times \be_{z} - h_{p}(\bm.\be_{y})\bm \times \be_{y} +v\bm \times (\bp \times \bm) +vr \bm \times \bp\},
\label{eq55}
\end{eqnarray}
\noindent
where the relative strength of the easy plane anisotropy $h_{p}=H_{p0}/H_{u0}$, using the notation of Eqs.(5), (6). The last two terms correspond to $\bTpar$, with strength parameter $v$, and $\bTper$, with strength parameter $rv$. Clearly the reduced bias $v$ is proportional to the actual bias $V_{b}$ and inversely proportional to the number of atomic planes in the switching magnet. Comparing Eqs.(47), (48), and (55) we see that $r=g_{\perp}/g_{\parallel}$. Thus if the scalar torques $T_{\parallel}$, $T_{\perp}$ defined after Eq.(50) have the same sign it follows that $r$ is positive. Conversely, $r$ is negative if the torques have opposite sign. We now linearize Eq.(51) about a steady-state solution $\bm=\bm _{0}$, which satisfies $\bG=0$, using the local coordinate axes shown in Fig.9. Thus
\begin{figure}[here]
\centerline{\epsfig{figure=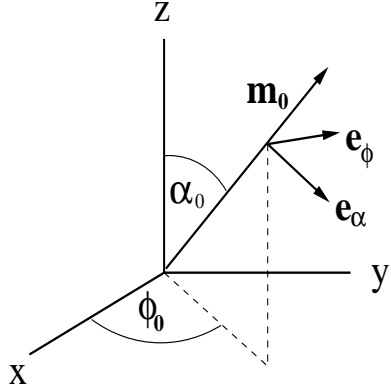,width=2in,height=2in,angle=0}}
\caption{\footnotesize Local coordinate axes for the deviation of the magnetization $\bm$ from its steady-state orientation $\bm_{0}$.} 
\end{figure}
\noindent  
\begin{eqnarray}
\bm = \bm _{0} + \xi \be_{\alpha} + \eta \be_{\phi}
\label{eq56}
\end{eqnarray}
\noindent
and the linearized Eq.(53) may be written in the form
\begin{eqnarray}
\frac{d\xi}{d\tau} = A\xi + B \eta; \; \frac{d\eta}{d\tau} = C\xi + D \eta,
\label{eq57}
\end{eqnarray}
\noindent
where $A$, $B$, $C$, $D$ are functions of $\alpha_{0}$, $\phi_{0}$, $\theta$ and the parameters $h_{p}$, $v$, $r$, and $\gamma$.
Following Sun \cite{sun}, we have introduced the natural dimensionless time variable $\tau = tH_{u0}/(1+\gamma ^{2})$. The conditions for the steady state to be stable are 
\begin{eqnarray}
F=A+D \le 0; \; G=AD-BC\ge 0
\label{eq58}
\end{eqnarray}
\noindent
excluding $F=G=0 $\cite{dynamical}. The damping parameter $\gamma $ appears in $G$ only as a factor $1+\gamma ^{2}$ and may be cancelled in the condition for stability $G\ge 0$. This condition becomes
\begin{eqnarray}
&  & Q^{2}v^{2}+(Qvr+\cos 2\alpha_{0})(Qvr+\ca)+h_{p}\{Qvr(1-3\sph \; \sa)+\cos 2\alpha_{0}(1-2\sa \;\sph)\} - \nonumber \\
&  & h_{p}^{2}\sa \; \sph(1-2\sph \; \sa)\ge 0,
\label{eq59}
\end{eqnarray}
\noindent
where $Q=\bp . \bm_{0}=\sin \theta \sin \alpha_{0} \cos \phi_{0} + \cos \theta \cos \alpha_{0}$.
Similarly, the condition $F\le0$ becomes
\begin{eqnarray}
-2v(1+\gamma r)Q - \gamma(\cos2\alpha_{0} + \ca) - \gamma h_{p}(1-3\sph \; \sa) \le 0.
\label{eq60}
\end{eqnarray}
\noindent
A number of general conclusions can be drawn from these inequalities. However, we first consider the special case when the magnetization of the polarizing magnet is in the direction of the uniaxial anisotropy axis of the switching magnet ($\theta =0$). In this case the equation $\bG=0$, with $\bG$ given by Eq. (55), shows immediately that possible steady states are given by $\alpha_{0}=0, \pi$, corresponding to the switching layer moment along the axis of the uniaxial anisotropy. These are the only solutions when $h_{p}=0$. However, in the presence of easy-plane anisotropy ($h_{p}\ne 0$) there are additional steady state directions of the switching layer moment given by $\sin 2\phi_{0} = 2v/h_{p}$, $\cos \alpha_{0} = -vr\cos\phi_{0}/(\cos\phi_{0}+v\sin \phi_{0})$.  
In the pure Slonczewski case of $\bTper =0$ ($r=0$) it follows that $\alpha_{0}=\pi/2$. The stability conditions (59) and (60) then reduce to $\cos 2\phi_{0}\le 0$ and $h_{p}(1-3\sin^{2}\phi_{0})\ge 1$ respectively. For practical biases $v/h_{p}\ll 1$ and the solution for $\phi_{0}$ satisfying the first stability condition is $\phi_{0}\approx \pi/2-v/h_{p}$. The second stability condition is then not satisfied. Thus for $\theta =0$, $r=0$ the only stable steady states of interest are given by $\alpha_{0}=0$ or $\pi$. It is easily seen that in this case the inequality (59) is always satisfied and the system becomes unstable when the left-hand side of Eq.(60) changes sign at $v_{c}=-(1+\frac{1}{2}h_{p})\gamma$, corresponding to Sun's \cite{sun} result in zero external field. This shows clearly that the criterion for magnetization switching derived by Slonczewski \cite {slon} and Sun \cite{sun} in their very special case is equivalent to the instability of a steady state in our approach.

In another special case when we have only out-of-plane torque $\bTper$, the switching criteria are clearly the same as in the Stoner-Wohlfarth field-switching theory since the torque $\bTper$ is equivalent to one arising from an effective field of magnitude $H_{u0}vr$ (see Eq.(55)). The stability criteria are then equivalent to the conditions for a minimum of an energy function whose gradient gives all the effective fields. This was previously recognized by Heide et al.  \cite{heide}. Obviously this energy does not involve $\gamma$ and the absence of $\gamma$ in both criteria is clearly seen when  in the second criterion (60) we take the limit $v\rightarrow 0, r\rightarrow \infty$ with the $\bTper$ parameter $vr$ finite. As soon as the in-plane torque parameter $v$ is nonzero, no energy function exists and the stability criterion (60) involves the damping parameter $\gamma$, showing its essentially dynamic nature. It is interesting that even when both spin-transfer torques exist, the in-plane torque drops out of Eq.(60) in the strong damping limit $\gamma \rightarrow \infty$ and we return to a Stoner-Wohlfarth situation.

In general, the system is neither in the Slonczewski-Sun nor Stoner-Wohlfarth limit and our general stability analysis based on the criteria (59) and (60) is required. We shall first apply it to discuss the stability of the steady-state paths shown in Fig.3 which correspond to the microscopic parameters (b) of Table I. In this case, the reduced parameters of the present section, which reproduce accurately the microscopically determined curves in Fig.3, are $h_{p}=0$ and $r=1$. The torque ratio $r=1$ is clear from the curves for case (b) in Fig. 4. The corresponding steady-state paths are shown in Fig.10 (a). In Fig. 10 (b) we plot the steady-state paths for the parameter set (d) of Table I, in which case $r=-1$ (see Fig. 4). In Fig. 3 we plotted $-V_{b}$ on the bias axis, where $V_{b}$ is in volts, in order to compare with the present reduced bias $v$ which is proportional to $eV_{b}$ and $e$ is negative. ($V_{b}$ and $v$ have the same sign when $T_{\parallel}<0$ and opposite sign when $T_{\parallel}>0$.) 
\begin{figure}[here]
\centerline{\epsfig{figure=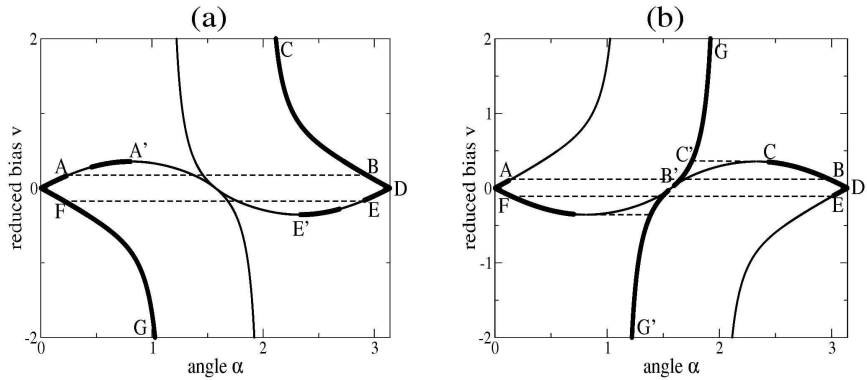,width=4.5in,height=2in,angle=0}}
\caption{\footnotesize Reduced bias $v$ required to stabilize the switching magnet moment  at an angle $\alpha$ on the universal path for $r=1$ (a) and $r=-1$ (b) and $h_{p}=0$. Bold lines correspond to stable steady states.} 
\end{figure}
\noindent  
We have chosen for presentation purposes a value $\gamma =0.05$ of the damping parameter which is somewhat larger than that suggested in  \cite{sun}. We recall that Figs.3 and 10 correspond to the situation when the moment of the polarizing magnet is at $\theta =2$ radians. We first discuss Fig. 10 (a). Initially with zero bias $v$ the moment of the switching magnet is in the direction of the uniaxial anisotropy field, i.e., $\alpha_{0}=0$. As $v$ increases positively, $\alpha_{0}$ increases until the stability criterion (60) ceases to be satisfied. In Fig.10 stable steady states are indicated by heavy lines  and unstable ones by thin lines. At this  value of $v$, the point A in Fig.10 (a), the system seeks another steady state which is stable at the same bias and  hence jumps to B. It is assumed that for finite $\gamma$ the system will home in on the stable state. Sun \cite{sun} showed how this happens dynamically for the special case of $\theta =0$. On further increase of bias the system proceeds to C and, on reducing the bias to zero, it moves to D where $\alpha=\pi$. The current-induced switching process is thus completed with the magnetization switched between the two stable zero-bias orientations ($\alpha=0$ and $\pi$) in the uniaxial anisotropy field. To reverse the process the bias is reversed and the system proceeds to E where it becomes unstable and jumps to F. Finally a further negative increase of $v$ takes the system to G and, on reducing the bias to zero, we return to O. If the resistance is calculated for each steady state, using our Keldysh formalism (or equivalently the Kubo formula) for charge current, the corresponding hysteresis loop of resistance versus bias can be plotted. We shall do this for later examples.

In the above example, the in-plane and out-of-plane torques are of equal strength and of the same sign ($r=1$). The instabilities at A and E are governed by the dynamical criterion (60). In the Stoner-Wohlfarth-like case discussed earlier the system remains stable, as bias is increased, up to the maximum at A'. The first criterion (59) determines the instability at this point and similarly at E'.

Fig. 10b shows the situation for the less usual case of negative $r$, in
particular $r=-1$ corresponding to parameter set (d). Starting at $\alpha_{0} = 0$, with bias increasing from zero, an instability occurs at A as before. However,
there are now two possible stable steady states to which the system might jump,
labelled by B' and B. We cannot say to which point the jump occurs without
following the detailed dynamics of the system with time-dependent solutions of
the Landau-Lifshitz equation. If the system jumps to B' further increase of bias
leads towards G where the moment of the switching magnet approaches alignment
with that of the polarizing magnet. However the bias is varied now, through
positive and negative values, the system remains on the stable steady state
branch GG' and no switching to the point D can occur. If, however, the system
jumps from A to B, the bias can be reduced to zero at D and a hysteresis loop
can be completed via E and F. On the other hand if on jumping to B the bias is
further increased, to reach the state C, a jump will occur to C' and the system
is again trapped on the branch GG'.  

The next example we consider employs a mapping of the fully realistic microscopic torques for Co/Cu(111) shown in Fig. 7b onto the macroscopic model. In this case the switching magnet consists of two atomic planes of cobalt ($M=2$). We recall that the nonmagnetic spacer consists of 20 atomic planes of Cu.
A new feature of this example is that we now introduce a strong easy-plane anisotropy with $h_{p}=100$. If we take $H_{u0}=1.86\times 10^{9}sec^{-1}$ corresponding to an uniaxial anisotropy field of about 0.01T, this value of $h_{p}$ corresponds to the shape anisotropy for a magnetization of $1.6\times10^{6}A/m$, similar to that of Co \cite{sun}. We take again $\theta =2$ radians and a realistic value of $\gamma =0.01$, which is in line with the value quoted by Sun \cite{sun}. The value of $r$ which gives a reasonable fit to the microscopic torques shown in Fig. 7b is $r=0.65$. We find that the strong easy-plane anisotropy forces the switching magnet moment to rotate in the ($x,z$) plane, which means that the universal paths in the ($\alpha, \phi$) plane are almost straight lines with $\phi=0$ or $\pi$. The plot of reduced bias $v$ against $\alpha $ is shown in Fig. 11 and again the heavy portions of the curves indicate stable steady states. The multiple loops of steady states in Fig. 11a are a new feature appearing for $h_{p}\ne0$. However, in this case,
\begin{figure}[here]
\centerline{\epsfig{figure=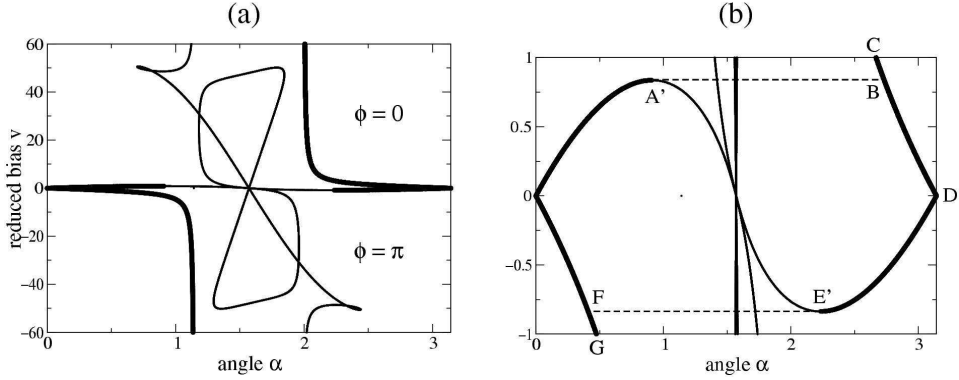,width=5in,height=2in,angle=0}}
\vspace{0.5cm}
\caption{\footnotesize Reduced bias $v$ required to stabilize the Co switching magnet moment at an angle $\alpha$, with $\theta =2$, and realistic anisotropy and damping parameters for Co/Cu/Co(111) with $M=2$. The value $\phi \approx 0$ is obtained for $v>0$ and $\phi \approx \pi$ is obtained for $v<0$.} 
\end{figure}
\noindent
they are all unstable. The important parts of the curves are shown on a larger scale in Fig.11b. This clearly resembles Fig. 10a and instabilities now occur at the extremal points A' and E' instead of A and E as in Fig.10a. 
In general, the point A lies further up the curve towards A' the larger the product $\gamma h_{p}$. This follows from Eq.(60) as long as the easy-plane anisotropy is strong enough for $\mid \sin \phi_{0} \sin \alpha_{0} \mid<1/\sqrt{3}$ to be satisfied.
This stabilizing effect of easy-plane anisotropy has the unwelcome consequence that the critical bias (current) for switching is strongly increased by such anisotropy. The corresponding hysteresis loop of resistance versus bias for $\theta=2$ radians is shown in Fig. 12a. We have also transferred the key points from Fig. 11b. 
In Fig. 12b we show the hysteresis loop for $\theta=3$ radians, which is close to $\theta=\pi$ assumed in previous treatments \cite{slon}, \cite{sun}. 
\vspace{0.2cm}
\begin{figure}[here]
\centerline{\epsfig{figure=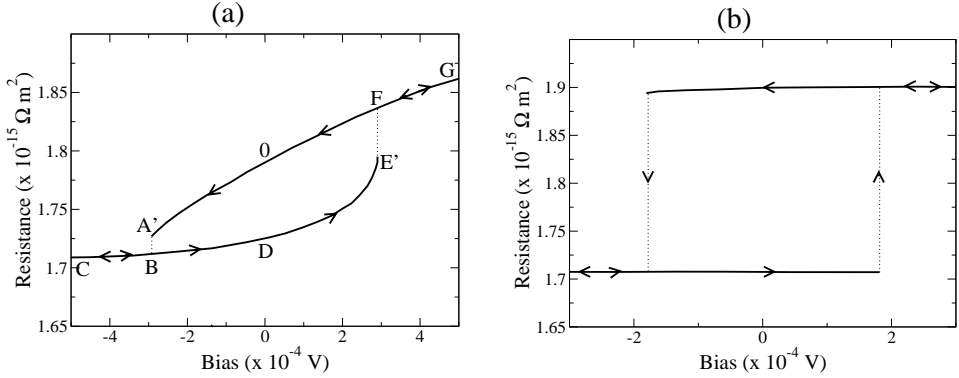,width=5in,height=2in,angle=0}}
\vspace{0.5cm}
\caption{\footnotesize Resistance of the Co/Cu/Co(111) junction as a function of applied bias with $M=2$ monolayers of Co in the switching magnet. (a) is for $\theta=2$ radians and (b) is for $\theta=3$ radians.} 
\end{figure}
\noindent  
It is rather interesting that the critical bias for switching is $\approx$ 0.2mV both for $\theta=2$ and $\theta=3$ radians. When this bias is converted to the current density using the calculated ballistic resistance of the junction, we find that the critical current for switching is $\approx 10^{7}A/cm^{2}$, which is in very good agreement with experiment \cite{pillar}. However, there is a qualitative difference between the cases $\theta=2$ and $\theta=3$ radians. For $\theta=2$ radians, switching is determined by the instability condition (59) which is independent of the damping parameter $\gamma$. This means that switching is of the Stoner-Wohlfarth type. On the other hand, for $\theta=3$ radians, we find that the instability is determined by Eq.(60), which means that switching is of the Slonczewski-Sun type.

The last example we consider employs again a mapping of the fully realistic microscopic torques for Co/Cu(111) shown in Fig. 7a onto the macroscopic model. In this case the number of atomic planes in the switching magnet is $M=1$. We use the same values of $h_{p}$, $\gamma$, $H_{u0}$, and $\theta$ as in the previous example. The best fit to the microscopic torques in Fig. 7a gives $r =-1.0$. This example, besides being again a realistic one, introduces the  feature of negative $r$ which we met in the single-orbital model with parameter set (d) (see Figs. 4 and 10b). The plot of reduced bias $v$ against $\alpha$ is shown in Fig. 13 with stable steady states indicated as before. The low-bias part of the curves is plotted in Fig. 13b on a larger scale.
\begin{figure}[here]
\centerline{\epsfig{figure=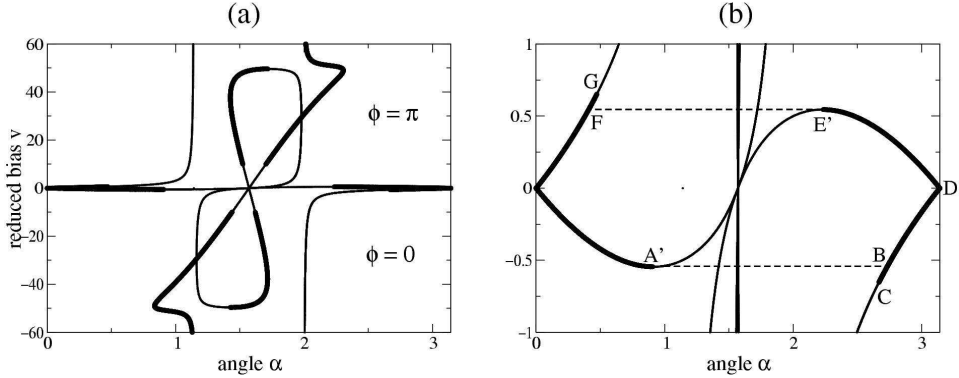,width=5in,height=2in,angle=0}}
\vspace{0.5cm}
\caption{\footnotesize Reduced bias $v$ required to stabilize the Co switching magnet moment at an angle $\alpha$, with $\theta =2$, and realistic anisotropy and damping parameters for Co/Cu/Co(111) with $M=1$. The value $\phi \approx 0$ is obtained for $v<0$ and $\phi \approx \pi$ is obtained for $v>0$.} 
\end{figure}
\noindent
The fact that $\mid r \mid$ is comparable for Figs. 11 and 13 (of the order of 1) but the sign is changed, leads to all the bias curves being essentially only reflected in the $\alpha$ axis. However, the change of sign of $r$ is seen to have a dramatic effect on the stability of the steady states.  
The most striking effect is the creation of a 'bias-gap' with no stable steady
states for values of reduced bias $\mid v\mid $ between about 0.7 and 10. For $r=-1$ the
bias-gap only exists in the presence of easy-plane anisotropy. In this
connection we may compare Fig.13, with parameters $r=-1$, $h_{p}=100$,        $\gamma=0.01$,
$\theta=2$ with Fig.10b correponding to $r=-1$, $h_{p}=0$, $\gamma=0.05$, $\theta=2$. (The
larger value of $\gamma$ for Fig.10b is not important; it was used to push the
point of instability A to larger bias and thus clarify the figure). Clearly
there is no bias-gap in Fig. 10b with $h_{p}=0$. Another effect of large 
$h_{p}$ is to
push the point of instability G in Fig. 13b to much larger bias than the
corresponding point in Fig.10b, even with a smaller damping parameter $\gamma$.
In fact, for the particular values of $\gamma$ and $h_{p}$ used for Fig. 13, G lies at
a larger bias than E'. The resultant hysteresis loop, shown in Fig.14a, is thus
executed in the same sense as that shown in Fig.12a. However in Fig. 14b,
corresponding to $\theta =3$ instead of 2, the sense is reversed.
\vspace{0.5cm}
\begin{figure}[here]
\centerline{\epsfig{figure=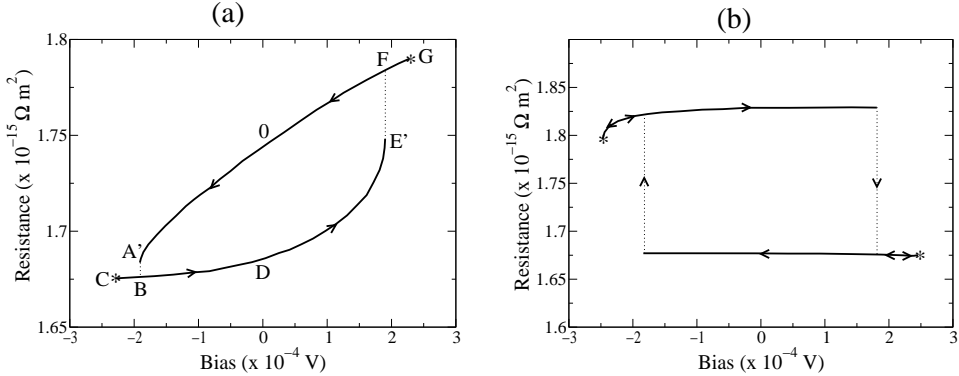,width=5in,height=2in,angle=0}}
\vspace{0.5cm}
\caption{\footnotesize Resistance of the Co/Cu/Co(111) junction as a function of applied current, with $M=1$ monolayer of Co in the switching magnet. (a) is for $\theta=2$ radians and (b) is for $\theta=3$ radians.} 
\end{figure}
\noindent
In Fig. 12 as we change bias from B to C or from F to G we are achieving saturation by aligning the switching magnet parallel or antiparallel to the polarizing magnet. However, in Fig. 14 as we increase bias from F the steady state becomes unstable at the point G, for a critical bias, but there is no stable state for the system to jump to with further increase of bias (see Fig. 13b). To emphasize this point, the points G and C are starred in Fig. 14a. Thus for bias larger than this critical one the system cannot home in on any stable state and the moment of the switching magnet remains perpetually in a time-dependent state. For much larger bias the system can home in onto the stable multiple loop states shown in Fig. 13a. Thus there is a range of bias where only the time-dependent state is possible. The bottom of the gap occurs at the bias point G in Fig. 14a. If the bias is then reduced below this value the system will home in on a stable steady state and the hysteresis loop can be completed. 

We investigated the critical negative value of $r$ at which the bias gap appears. For the parameters used above, the gap is not present for $r=-0.05$ but is already well established for $r=-0.1$. Thus when in-plane and out-of-plane spin-transfer torques have opposite sign, and easy plane anisotropy is large, only a small out-of-plane torque is required to produce this unusual bias gap behavior. Since out-of-plane torque corresponds to an effective field, we believe that this behavior is closely related to the time-dependent motion of the moment of the switching magnet which is observed in a sufficiently large applied magnetic field \cite{kiselev}. This alternative mechanism for time-dependent motion of the switching magnet moment is under investigation.

Even in the absence of easy plane anisotropy, but with large negative  $r$, we have found a critical bias above which only time-dependent solutions exist. However, for small negative $r$ ($\mid r \mid \ll 1$) normal solutions can occur (no bias gap). For intermediate values of $r$, as discussed for $r=-1$ with reference to Fig. 10b, switching may or may not occur. Furthermore, in the case of pure in-plane torque ($r=0$) and no easy plane anisotropy we find straightforward switching of Slonczewski-Sun type for $\mid \cos \theta \mid>1/3$, while for $\mid \cos \theta \mid<1/3$ a more complicated hysteresis loop is found.

\section{Conclusions}

Our principal result is that spin-transfer torques responsible for current-induced switching of magnetization can be calculated quantitatively for real systems such as Co/Cu/Co junction in the ballistic regime using nonequilibrium Keldysh formalism. In fact, we argue that the spin-transfer torque can be calculated selfconsistently from first principles only in a steady state (switching magnet magnetization does not move), and this is precisely what the Keldysh formalism is designed for. In the small-bias (linear-response) regime higher-order (many-body) effects can be neglected and our results for the spin-transfer torque are, therefore, quite rigorous. Keldysh formalism provides an explicit formula for the local spin current between any two atomic planes of the junction in terms of  one-electron surface Green functions for the cut junction. The surface Green functions are readily available and we calculate them using a tight-binding Hamiltonian with parameters determined from a fit to an {\it ab initio} band structure. With the exception of Slonczewski's parabolic band calculation \cite{slon}, our Keldysh formulation is the only theory that yields the local spin current taking into into account rigorously contributions from all the parts of the junction. As the following argument demonstrates, previous theories \cite{stiles}, \cite{fert},  which consider only scattering of spin-polarized electrons incident from the spacer on the spacer/switching magnet interface are incomplete. When the particle current flows from the polarizing magnet toward the switching magnet it is clear that a spin-transfer torque acts on the switching magnet. However, when the polarity of the applied bias is reversed, the current incident from the right lead on the switching magnet is unpolarized and, therefore, has no effect on it. On the other hand, it is well known experimentally \cite{pillar} that changing the polarity of the bias reverses the direction of the spin-transfer torque (the magnitude remains the same). This cannot be explained within a theory that treats only the spacer/switching magnet interface. The ingredient that is missing is strong reflection of electrons from the polarizing magnet which results in a spin polarization of the reflected electrons. Clearly only such reflected spin current flowing in the direction opposite to that of the particle current can exert torque on the switching magnet. It follows that multiple reflections of electrons from the polarizing and switching magnet are an essential feature of the problem. They are treated rigorously to all orders in our theory. The fundamental experimental fact that the spin-transfer torque acting on the switching magnet is proportional to the applied bias is obtained naturally in our theory since the spin current anywhere in the junction, given by Eq.(25), is proportional to the difference between the Fermi functions for the left and right halves of the cut junction, i.e., proportional to the bias in the low-bias (linear-response) limit. These arguments indicate that selfconsistent treatment of the whole junction is crucial for correct understanding of current-induced switching of magnetization. 

The spin-transfer torque calculated from our Keldysh formalism has two components, one with the torque vector $\bTpar $ in the plane containing the magnetizations of the two magnetic layers ('in-plane' torque) and another with the torque vector $\bTper $ perpendicular to this plane ('out-of-plane' torque). It is generally believed that the effective field-like component $\bTper $ is always small. We find that this is not the case and our calculations show that, in general, both the in-plane and out-of-plane components tend to finite values independent of the spacer thickness in the limit of a thick spacer. However, it is true that $\bTper $ is strictly zero in the limit of an infinite exchange splitting between the majority and minority-spin bands in both ferromagnets, and this is the case considered initially by Slonczewski \cite{slon}. In the realistic case of a finite exchange splitting, $\bTper $ is nonzero and can be comparable with $\bTpar $. The only other general case when  $\bTper $ can be small occurs for a junction with reflection symmetry about a plane at the center of the spacer. Hence to observe an effect of $\bTper $ one needs to break the reflection symmetry of the junction. For a junction with polarizing and switching magnets made of the same material, this is achieved by making the switching magnet thinner than the polarizing magnet, and a strongest effect is found for a switching magnet only a few atomic planes thick. Our calculations show that $T_{\perp} $ and $T_{\parallel} $ are comparable for a Co/Cu/Co(111) junction when the switching Co layer is one or two atomic planes thick. Nevertheless, for a good epitaxial junction (ballistic limit), we find that $T_{\perp} $ is $\approx$ 27\% of $T_{\parallel} $ even for a switching Co magnet as thick as ten atomic planes. An alternative way to break the symmetry is to use a junction with polarizing and switching magnets made of different materials. 

Another result we wish to highlight is that, depending on material parameters of the junction, the relative sign of $T_{\perp}$ and $T_{\parallel}$ can be negative as well as positive. For example, $T_{\perp}/T_{\parallel}<0$ for Co/Cu/Co(111) with a switching Co magnet of one atomic plane and $T_{\perp}/T_{\parallel}>0$ for two atomic planes of Co. The negative sign of the ratio $T_{\perp}/T_{\parallel}$ has a profound effect on the stability of steady states and, hence, on the nature of current-induced switching. 

Finally, to determine the critical currents for switching and to investigate the effect of $\bTper $, we have used the microscopically calculated spin-transfer torques as an input into the phenomenological Landau-Lifshitz equation with Gilbert damping. Our general philosophy is that all steady states can be calculated from first principles and loss of their stability, determined from the Landau-Lifshitz equation, corresponds to switching. This holds provided there is another stable steady state at the same current density the system can switch into. We showed that our criterion for instability of the steady state  leads to the same critical current for switching as that obtained earlier by Sun \cite{sun} in the special case of $\bTper =0$ and for the initial angle $\theta$ between the polarizing and switching magnet moments equal to $0$ or $\pi$. However, we find that qualitatively different switching scenarios can occur when $T_{\perp}/T_{\parallel} \ne 0$, $\theta \ne 0$, and in the presence of an easy-plane (shape) anisotropy. In particular, when the easy-plane anisotropy is strong, even a relatively small $T_{\perp} $ (5-10 \% of $T_{\parallel}$) has a strong effect on switching. In the absence of an applied magnetic field, we find that an ordinary hysteresis loop is the only possible switching scenario when $T_{\perp}/T_{\parallel}>0$. However, for $T_{\perp}/T_{\parallel}<0$, a normal hysterestic switching occurs only at relatively low current densities. When the current exceeds a critical value, there are no stable steady states and the system thus remains permanently in a time dependent state. This is analogous to the observed precession of the switching magnet magnetization caused by a DC current in the presence of an applied magnetic field \cite{kiselev}. In our case, the effective field-like term $\bTper $, which causes this behavior, is proportional to the DC current and, hence, complete loss of stability of the steady state occurs only when this term is large enough, i.e., when the DC current is above a critical value.

Our calculations for Co/Cu/Co(111) show that the critical current for switching in the hysteretic regime is $\approx 10^{7}A/cm^{2}$ both for Co switching magnets of one and two atomic planes. This is in good agreement with experiment \cite{pillar}. We recall that the critical current for switching is obtained using the spin transfer torques for a fully realistic Co/Cu/Co(111) junction and assuming a uniaxial anisotropy of $0.01$T and Gilbert damping $\gamma =0.01$. This is in line with the values of the uniaxial anisotropy and Gilbert damping quoted by Sun \cite{sun}.

We conclude by stressing that all the specific results we have obtained are strictly  valid for a perfect epitaxial junction, i.e., in the ballistic limit. However, the Keldysh formalism we have described is valid also in the diffusive limit. Generalization to the diffusive limit is, in principle, straightforward. For example, one could introduce random impurities in the lateral supercell geometry, determine the one-electron surface Green functions in this geometry  and then perform configuration averaging of the spin current.  

\section{Acknowledgments}

The support of the Engineering and Physical Sciences Research Council (EPSRC UK) is gratefully acknowledged. We also acknowledge stimulating discussions with J.C. Slonczewski, R.A. Buhrman, and J.Z. Sun.


\begin{references}
\bibitem{slon} 
J.C. Slonczewski, J. Magn. Magn. Mater. {\bf 159}, L1 (1996);  J. Magn. Magn. Mater. {\bf 195}, L261 (1999); J. Magn. Magn. Mater. {\bf 247}, 324 (2002).
\bibitem{berger}
L. Berger, J. Appl. Phys. {\bf 49}, 2156 (1978); Phys. Rev. B {\bf 54}, 9353 (1996).
\bibitem{pillar}
M. Tsoi, A.G.M. Jansen, J. Bass, W.C. Chiang, M. Seck, V. Tsoi, and P.Wyder, Phys. Rev. Lett., {\bf 80}, 4281 (1998); 
M. Tsoi, A.G.M. Jansen, J. Bass, W.C. Chiang, V. Tsoi, and P. Wyder, Nature {\bf 406}, 46 (2000);
E.B. Myers, D.C. Ralph, J.A. Katine, R.N. Louie, and R.A. Buhrman, Science {\bf 285}, 867 (1999); J.A. Katine, F.J. Albert, R.A. Buhrman, E.B. Meyers, and D.C. Ralph, Phys. Rev. Lett. {\bf 84}, 3149 (2000); F.J. Albert, J.A. Katine, R.A. Buhrman, and D.C. Ralph, Appl. Phys. Lett. {\bf 77}, 3809 (2000); J. Grollier, V. Cros, A. Hamzic, J.M. George, H. Jaffres, A. Fert, G. Faini, J. Ben Youssef, and H. Legall, Appl. Phys. Lett. {\bf 78}, 3663 (2001).
\bibitem{stoner}
E.C. Stoner and E.P. Wohlfarth, Phil. Trans. Roy. Soc. A {\bf 240}, 599 (1948).
\bibitem{sun}
J.Z. Sun, Phys. Rev. B {\bf 62}, 570 (2000).
\bibitem{kiselev}
S.I. Kiselev, J.C. Sankey, I.N. Krivorotov, N.C. Emley, R.J. Schoelkopf, R.A. Buhrman, and D.C. Ralph, Nature {425}, 380 (2003).
\bibitem{grollier}
J. Grollier, V. Cros, H. Jaffres, A. Hamzic,J.M. George, G. Faini, J. Ben Youssef, H. Le Gall, and A. Fert, Phys. Rev. B {\bf 67}, 174402 (2003).
\bibitem{stiles}
M.D. Stiles and A. Zangwill, Phys. Rev. B {\bf 65}, 014407 (2002); M.D. Stiles, Jiang Xiao, and A. Zangwill, Phys. Rev. B {\bf 69}, 054408 (2004).
\bibitem{keldysh}
L.V. Keldysh, Soviet Physics JETP {\bf 20}, 1018 (1965).
\bibitem{caroli}
C. Caroli, R. Combescot, P. Nozieres, and D. Saint-James, J. Phys. C {\bf4}, 916 (1971).
\bibitem{edwards}
D.M. Edwards in: Exotic states in quantum nanostructures, ed. by S. Sarkar, Kluwer Academic Press (2002).
\bibitem{prb97}
J. Mathon, Murielle Villeret, A. Umerski, R.B. Muniz, J. d'Albuqerque e Castro, and D.M. Edwards, Phys, Rev. B {\bf 56}, 11797 (1997).
\bibitem{book}
G.D. Mahan, Many Particle Physics, 2nd Ed., Plenum Press, New York (1990).
\bibitem{exchcp}
D.M. Edwards, A.M. Robinson, and J. Mathon, J. Magn. Magn. Mater. {\bf 140-144}, 517 (1995).
\bibitem{fert}
S. Zhang, P.M. Levy, and A. Fert, Phys. Rev. Lett. {\bf 88}, 236601 (2002).
\bibitem{heide}
C. Heide, P.E. Zilberman, and R.J. Elliott, Phys. Rev. B {\bf 63}, 064424 (2001).
\bibitem{dynamical}
D.W. Jordan and P. Smith, Nonlinear Ordinary Differential Equations, Clarendon Press, Oxford (1977). 





\end{references}
\end{document}